\RequirePackage{ifpdf}
\documentclass{article}
\usepackage{jheppub}
\hypersetup{
	colorlinks=true,
	linktocpage=true,
	linkcolor=blueviolet,
	filecolor=blueviolet,
	urlcolor=blueviolet,
	citecolor = blueviolet,
	}
\usepackage{xcolor}
\usepackage{amsmath}
\usepackage{mathtools}
\usepackage{epsfig}
\usepackage{environ}
\usepackage{comment}
\usepackage{soul}
\pdfoutput=1

\usepackage[normalem]{ulem}

\newcommand{\fixme}[1]{{\bf {\color{blue}[#1]}}}

\newcommand{\roughly}[1]{\mathrel{\raise.3ex\hbox{$#1$\kern-0.85em
\lower1ex\hbox{$\sim$}}}}

\newcommand{\be}{\begin{equation}}
\newcommand{\bee}{\begin{equation}}
\newcommand{\ee}{\end{equation}}
\newcommand{\beea}{\begin{eqnarray}}
\newcommand{\eea}{\end{eqnarray}}
\newcommand{\bea}{\begin{eqnarray}}
\newcommand{\ea}{\end{eqnarray}}
\newcommand{\ba}{\begin{eqnarray}}
\newcommand{\eq}{\begin{equation}}
\newcommand{\eeq}{\end{equation}}
\newcommand{\eqa}{\begin{eqnarray}}
\newcommand{\eeqa}{\end{eqnarray}}

\definecolor{blueviolet}{RGB}{138,43,226}

 \NewEnviron{ignore}{}{}  

\def\nott#1{\setbox0=\hbox{$#1$}                
   \dimen0=\wd0                                 
   \setbox1=\hbox{/} \dimen1=\wd1               
   \ifdim\dimen0>\dimen1                        
      \rlap{\hbox to \dimen0{\hfil/\hfil}}      
      #1                                        
   \else                                        
      \rlap{\hbox to \dimen1{\hfil$#1$\hfil}}   
      /                                         
   \fi}                                         %

\def\uxsl{\hbox{/\kern-.4000em$u$}}
\def\uxslsm{\hbox{\smaller/\kern-.5600em$u$}}
\def\pxpsl{\hbox{/\kern-.5000em$p$}}
\def\epssl{\hbox{/\kern-.5600em$\epsilon$}}
\def\delsl{\hbox{/\kern-.7000em$\nabla$}}
\def\lxpsl{\hbox{/\kern-.5600em$l$}}
\def\kxpsl{\hbox{/\kern-.5600em$k$}}
\def\qxpsl{\hbox{/\kern-.3900em$q$}}

\def\UV{{\scriptscriptstyle U\hbox{\kern-0.1em}V}}
\def\PPN{{\scriptscriptstyle P\hbox{\kern-0.1em}P\hbox{\kern-0.1em}N}}
\def\MN{{\scriptscriptstyle M\hbox{\kern-0.1em}N}}
\def\MNP{{\scriptscriptstyle M\hbox{\kern-0.1em}N\hbox{\kern-0.1em}P}}
\def\KK{{\scriptscriptstyle K\hbox{\kern-0.1em}K}}
\def\SM{{\scriptscriptstyle S\hbox{\kern-0.1em}M}}
\def\EH{{\scriptscriptstyle E\hbox{\kern-0.1em}H}}

\def\QCD{{\scriptscriptstyle Q\hbox{\kern-0.1em}C\hbox{\kern-0.1em}D}}
\def\IR{{\scriptscriptstyle I\hbox{\kern-0.1em}R}}
\def\TEV{{\scriptscriptstyle T\hbox{\kern-0.1em}E\hbox{\kern-0.1em}V}}

\def\UV{{\scriptscriptstyle U\hbox{\kern-0.1em}V}}
\def\PPN{{\scriptscriptstyle P\hbox{\kern-0.1em}P\hbox{\kern-0.1em}N}}
\def\MN{{\scriptscriptstyle M\hbox{\kern-0.1em}N}}
\def\MNP{{\scriptscriptstyle M\hbox{\kern-0.1em}N\hbox{\kern-0.1em}P}}
\def\KK{{\scriptscriptstyle K\hbox{\kern-0.1em}K}}
\def\SM{{\scriptscriptstyle S\hbox{\kern-0.1em}M}}
\def\EH{{\scriptscriptstyle E\hbox{\kern-0.1em}H}}

\def\QCD{{\scriptscriptstyle Q\hbox{\kern-0.1em}C\hbox{\kern-0.1em}D}}
\def\IR{{\scriptscriptstyle I\hbox{\kern-0.1em}R}}
\def\TEV{{\scriptscriptstyle T\hbox{\kern-0.1em}E\hbox{\kern-0.1em}V}}
\def\aff{{a\hbox{\kern-0.1em}f\hbox{\kern-0.1em}f}}

\usepackage[numbers,sort&compress]{natbib}

\title{Universal Confining Strings:\\ From Compact QED to the Hadron Spectrum}
\author[a]{M.C. Diamantini,}

\author[b,c]{F. Quevedo,}

\author[b,d]{C.A. Trugenberger,}

\author[b]{L. Zapata}

\affiliation[a]{NiPS Laboratory, INFN and Dipartimento di Fisica e Geologia, \\ University of Perugia, via A. Pascoli, I-06100 Perugia, Italy.}
\affiliation[b]{Division of Science, New York University Abu Dhabi, Abu Dhabi, United Arab Emirates.}
\affiliation[c]{DAMTP University of Cambridge, Wilberforce Road, Cambridge, CB3 0WA, UK.}
\affiliation[d]{SwissScientific Technologies SA, rue du Rhone 59, CH-1204 Geneva, Switzerland.}

\emailAdd{cristina.diamantini@pg.infn.it}
\emailAdd{fq201@cam.ac.uk}
\emailAdd{ca.trugenberger@bluewin.ch}
\emailAdd{lfz2012@nyu.edu}

\date{today}

\abstract{We investigate the description of quark confinement in terms of confining strings or
flux tubes. We show that compact QED with a topological $\theta$-term, in the dyon condensation
phase, is described by a {\it massive} two-form field $B_{\mu\nu}$ that gives rise to a string theory with an IR  Brazovskii-Lifshitz fixed point at strong coupling, reached along a double-scaling renormalization trajectory on which the cutoff can be removed. This corresponds
to a UV-complete quantum string in (3+1) dimensions, representing the dual of asymptotic freedom in the UV.
Contrary to critical strings, which correspond to trivial Gaussian fixed points, this string is stabilized by a finite thickness, determined by the mass of the $B_{\mu\nu}$ field, instead of living in a higher-dimensional space. Interestingly, it contains a world-sheet excitation, in addition to the Nambu-Goto phonons, whose nature depends on the sign of the string stiffness: for the negative stiffness favoured by lattice data it appears as a broad, overdamped mode rather than a sharp resonance, consistent with the broad phase shift seen in lattice $SU(N)$ scattering data. On the other hand,
we determine the confining potential and show that it reproduces a generalized Arvis potential $V(L)= aL\sqrt{1-c/L^2}$ with running parameters $a(L), c(L)$. With this, we compute the mass difference
ratios for the heaviest quarkonium and find 2.5\% agreement with experiment already at the infrared fixed point, from a two-parameter fit of both the reduced quark mass and a boundary-condition parameter, to this fixed functional form. We also compute the intercept of Regge trajectories and find that the thickness of Brazovskii-Lifshitz strings tends to increase it from the Nambu-Goto value $\alpha_0= 1/12$. Overall, our findings strongly support Polyakov’s longstanding
conjecture on universality of confining gauge theories in the IR.
}

\begin{document}
\maketitle

\section{Introduction}
Understanding the mechanism of confinement in the strong interaction remains one of the central open problems of the Standard Model of particle physics. Over the past decades, using theoretical insights as well as lattice gauge theory calculations, a coherent theoretical picture has emerged, identifying the essential ingredients that such a mechanism is expected to involve \cite{Greensite2020, Gross:2022hyw}. A key element of this picture is the condensation of magnetic monopoles, which realises a dual analogue of the Higgs mechanism familiar from superconductivity \cite{tHooft:1977nqb, tHooft:1999cgx}. This leads to the electric-magnetic dual of the Meissner effect, whereby electric, rather than magnetic, flux is expelled from the vacuum and confined into flux tubes \cite{Mandelstam:1974pi, DiGiacomo:1999yas,DiGiacomo:1999fb,Carmona:2001ja}. These flux tubes are expected to manifest themselves as confining strings connecting, for instance, quark-antiquark pairs, thereby preventing the existence of free coloured particles at low energies.

Importantly, this feature has been observed in lattice gauge theory simulations of different systems \cite{Juge:2002br,HariDass:2007tx,Athenodorou:2011rx} shedding light on the structure of confining flux tubes \cite{Luscher:1980iy,Caselle:2016mqu}. Note also that confinement is not an exclusive feature of non-Abelian gauge theories, it has been also observed in abelian gauge theories, both in lattice simulations and in condensed matter systems. Indeed, Abelian confinement of Cooper pairs in the superinsulating state \cite{Diamantini:1995yb,Diamantini:2018mjg} has been recently detected in many materials (for a review see \cite{diamantini2023superinsulation}). Furthermore, Polyakov's proof of confinement in 3D was precisely for a compact $U(1)$ \cite{Polyakov:1975rs} (see \cite{Aharony:2024ctf} for a recent study of the string dual of this $3D$ model). This shows that the non-Abelian character of gauge fields is not the main driver of confinement. Instead, compactness of the gauge group may be the essential phenomenological ingredient.

Indeed we know that  QCD provides a meaningful description of strong interactions only at short scales, where asymptotic freedom renders the coupling weak and the theory perturbative.  At large scales, however, the coupling becomes strong and color is confined, leading to hadrons as the only observable states. In this confinement regime one expects strings, formed by color flux tubes to correctly describe the strong interactions.
While the potential role of strings in understanding confinement is intuitively appealing and widely accepted, a concrete and fully developed implementation remains elusive (see \cite{Aharony:2013ipa,Dubovsky:2012sh,Brandt:2016xsp} and reference therein for reviews of recent work). While the Nambu-Goto string captures correctly the linear potential and its universal L\"uscher correction \cite{Luscher:1980fr, Luscher:2002qv}, it cannot be the correct string theory \cite{Aharony:2010db,Caselle:2024zoh}, since it requires the  wrong spacetime dimensionality and fails to reproduce other key features of hadronic physics, such as particle masses and the intercept of Regge trajectories.

For this reason, in the prevailing view, consistent quantum bosonic strings in (3+1) dimensions can exist only as effective strings \cite{Polchinski:1991ax, Aharony:2009gg, Aharony:2010cx}. On the one hand, the holographic approach offers an effective description of confinement through a top-down approach \cite{Witten:1998zw} relaying on the AdS/CFT bulk-boundary correspondence \cite{Maldacena:1997re, Gubser:1998bc, Witten:1998qj} to compute observables in the gauge theory with a dual string theory (see \cite{Gabai:2025hwf, Gabai:2026myo} for recent work). On the other hand, an alternative approach is to write down all possible
corrections consistent with symmetries to the Nambu-Goto string \cite{Aharony:2013ipa,Dubovsky:2012sh} and compare with simulations of confining gauge theories \cite{Brandt:2016xsp, Teper:2009uf} or use bootstrap techniques to put rigorous bounds on observables \cite{Paulos:2016but,EliasMiro:2019kyf,EliasMiro:2021nul}. This approach also requires a cutoff in (3+1) dimensions.

Even though the effective approach has led to several insights into the properties that any string-like description of flux tubes must possess, there remains considerable skepticism regarding the existence of a UV-complete string theory beyond the critical dimension. The purpose of this paper is to explore this possibility.

Consequently, we follow an alternative approach, that of looking for a UV-complete string theory directly in (3+1) dimensions. If such a UV completion indeed exists, the IR fixed point defined by such a model may characterize an entire universality class of confining theories.
Duality suggests that strings are dual mirrors to gauge fields, especially for large colour limit. So if duality is realized, there should exist an infrared string fixed point at strong gauge coupling. Such a string fixed point should be universal and independent of colour, representing a genuine infrared string theory of confinement rather than a microscopic property of the gauge fields themselves, a dual mirror to asymptotic freedom. As emphasized by Polyakov \cite{Polyakov:1997tj}, if such a string fixed point can be found in any confining gauge theory, due to universality it should be the correct dual theory of QCD. At large scales, colour should play essentially no role.  In the holographic approach, it is believed that this fundamental string theory lives in a higher-dimensional spacetime. But there is no reason to assume that this must be true other than a UV-complete string theory in (3+1) dimensions is sometimes thought not to exist. The holographic approach, in addition, rests on several assumptions that question the validity of concrete top-down constructions.

Contrary to a widespread belief, however, a UV-complete string theory with long distance Nambu-Goto behaviour does exist in (3+1) dimensions \cite{Diamantini:1998zu, Diamantini:1999uj}. It is stabilized by a thickness emerging from dimensional transmutation, instead of 26 or 10 dimensions, and it realizes a Brazovskii-Lifshitz fixed point \cite{lifshitz1942theory, Hornreich:1975zz, brazovskii1975phase} where the cutoff can be removed. This, of course, makes Polyakov's conjecture even more appealing since only established (3+1)-dimensional field theory is needed to realize it.

In this work we thus take Polyakov’s proposal seriously and analyze compact QED in (3+1) dimensions as a paradigmatic example of a confining gauge theory in the strong-coupling regime, one that finds a direct physical realization in superinsulators (see \cite{diamantini2023superinsulation}). It is known that compact QED in the monopole confinement phase gives rise to  a massive antisymmetric tensor $B_{\mu\nu} $ field theory which, after proper integration of the massive field, admits a dual representation as a string theory \cite{Quevedo:1996uu,Polyakov:1996nc,Diamantini:1996vf}. Here we show that, when a $\theta$-term is added to the action (and the condensate is thus dyonic) this string formulation becomes exact at all scales and reduces precisely to the previously mentioned UV-complete Brazovskii-Lifshitz string. In presence of a $\theta$-term the cutoff can be removed and the model flows to the Brazovskii-Lifshitz point describing a string with a finite thickness. The thickness arises from the mass of the $B_{\mu\nu}$ field and is all that remains of the original gauge fields. Strings much longer than their thickness approach $c=1$ (per transverse coordinate) Nambu-Goto behaviour.

We quantize the BL string non-perturbatively in the $1/D$ expansion following the classic paper \cite{Alvarez:1981kc}. This permits us to derive the confining quark potential at the fixed point and to analyze the deviations from Nambu-Goto behaviour \cite{Luscher:1980fr,Alvarez:1981kc} at leading order in $1/D$. These can be expressed in the form of a generalized Arvis potential \cite{Alvarez:1981kc, Arvis:1983fp} $V(L)= a(L)L\sqrt{1-2b(L)/a(L)L^2}$ with running coefficients $a(L)$ and $b(L)$. As usual, there is a linear regime at large distances, while the potential drops strongly at shorter distances. The string is stable for sizes longer than 5.6 times its thickness. This critical scale plays the dual role to $\Lambda_{QCD}$ in asymptotically free gauge theories. For shorter strings one has to capture the appearance of non-string degrees of freedom on short scales by considering further terms in the $1/D$ expansion, exactly as the usual perturbative expansion of gauge theories breaks down at scales larger than $\Lambda_{\rm QCD}$.

To test the shape of the derived potential, we solve the Schr\"odinger equation for the heavy radial $\Upsilon(nS)$ states and we compare the ratios of eigenvalue differences to the corresponding ratios of $\Upsilon(nS)$ mass differences. Our prediction is in 2.5\% agreement already at the fixed point, without including any of the relevant perturbations of the Brazovskii-Lifshitz string. See section \ref{masssp}.

Furthermore, we find from this analysis that the Regge intercept $\alpha_0 (L)$ is also a function of the string length. At large $L$ it approaches the Nambu-Goto value 1/12 but for shorter scales it is enhanced to higher values.

Finally, let us mention that the Brazovskii-Lifshitz string automatically contains a world-sheet excitation in addition to the usual Nambu-Goto phonons, whose character depends on the sign of the stiffness $S$: a sharp resonance for $S>0$, and a broad, overdamped mode for the $S<0$ branch favoured by the lattice data discussed below. This excitation represents a breathing mode of the thick string and offers an alternative to the later introduced worldsheet axion \cite{Athenodorou:2024loq,Dubovsky:2015zey, Dubovsky:2013gi} to match lattice $SU(N)$ data. See section \ref{spectrums} for a more detailed discussion.

Overall, our results lend strong support to Polyakov’s universality conjecture, with the Brazovskii-Lifshitz IR fixed point representing the dual to the UV asymptotic freedom of gauge fields.

The rest of this work is structured as follows. We begin in section \ref{sec2} introducing the Brazovskii-Lifshitz string using heuristic arguments. We turn to compact $U(1)$ gauge theory in section \ref{sec3} where monopole configurations play a central role. The condensation of these monopoles induces a $4D$ field theory for a massive $B_{\mu\nu}$ field. This field has a natural coupling to string sources, and thus we proceed to integrate out the massive two-form field. In section \ref{sec4} we argue that the resulting action corresponds to a confining string that exhibits a Brazovskii-Lifshitz infrared fixed point at strong gauge coupling. The resulting model is further studied in the saddle-point approximation in section \ref{sec5}, where we comment on the string spectrum, compute the string potential, string tension and intercept. From this potential we extract the mass-difference ratios of the heaviest quark states. Finally, we present our conclusions in section \ref{sec6}.

\section{Brazovskii-Lifshitz Strings in (3+1) Dimensions}\label{sec2}

In this section we will introduce the Brazovskii-Lifshitz string. We start reviewing the standard Polyakov bosonic string, emphasising that it corresponds to a Gaussian fixed point and anomaly cancellation imposes $D=26$. The simplest deformation for each worldsheet coordinate corresponds to add a mass term  which unfortunately breaks the shift symmetry, making it unsuitable to represent a target space coordinate.   However, by going to a dual frame we show that this model can be written entirely using derivative operators of a dual scalar field. Because this dual representation preserves shift symmetry these dual fields can consistently represent the worldsheet embedding coordinates. The stability of this higher-derivative worldsheet theory is the focus of this and subsequent sections.

\subsection{Thick string actions}

It is generally believed that no UV-complete quantum strings can exist in (3+1) dimensions\footnote{Note however that the holographic correspondence has allowed to approach confining strings directly from the UV complete higher dimensional bulk string  theory, see  \cite{Aharony:2009gg,Teper:2009uf,Sonnenschein:2016pim} for reviews and reference therein.}. This situation, however reflects the traditional exclusive focus on Gaussian fixed points of the world-sheet theory for very good reasons. In the usual Polyakov formulation of string theory (for a review see \cite{polchinski2005string}), the (Euclidean) world-sheet action in flat target space spanned by $D$ coordinate fields $x_{\mu}$ is
\begin{equation}
S = \frac{1}{ 4\pi \alpha^{\prime}}\int d^2{\sigma} \sqrt{g} \ g^{ab}
\partial_a x_{\mu }
\partial_b x^{\mu } \ ,
\label{polac}
\end{equation}
where $\alpha'$ is the universal Regge slope, also related to the inverse string tension, $a,b, \cdots = 0,1$ are indices of the world-sheet coordinates while $\mu, \nu, \cdots$ are indices for target spacetime coordinates. Using diffeomorphism invariance one can first bring the world-sheet metric into conformally flat form. Using further Weyl invariance under reparametrizations of the metric by an overall factor one can obtain the flat metric $g_{ab} = \eta_{ab} = {\rm diag}(1,1)$. This gives the conformal gauge form
\begin{equation}
S = \frac{1}{ 4\pi \alpha^{\prime}}\int d^2\sigma \
\partial_a x_{\mu }
\partial_a x^{\mu } \ .
\label{cg}
\end{equation}
which, at least locally, is a classically well-defined theory. We already note that the steps leading to \eqref{cg} considerably simplify  quantizing \eqref{polac}. But conformal anomaly cancellation forces $D=26$. We then need an alternative string action to describe confining strings in $D=3+1$ dimensions. Since the flux tubes have tension and thickness the corresponding confining string may have tension but also another scale to give rise to thickness.

 The simplest perturbation away from the Gaussian fixed point is a mass term, which would be described by a Lagrangian
\begin{equation}
{\cal L} = \frac{1}{2} (\partial_{\mu} \varphi )^2 + \frac{1}{2} m^2\varphi^2 \ .
\label{sg}
\end{equation}
This, however, is not an allowed perturbation for a string interpretation, since it breaks translation invariance in target space. We cannot identify $\varphi$ with the embedding coordinates of \eqref{cg}.

On the other hand, a less studied massive model is the high-derivative scalar field theory known as the (isotropic) Brazovskii-Lifshitz model \cite{lifshitz1942theory, Hornreich:1975zz, brazovskii1975phase}, defined by a Lagrangian
\begin{equation}
{\cal L} = \frac{m^2}{ 2} (\partial_{\mu} \chi )^2\ \, + \frac{1} {2} (\partial_{\mu}^2 \chi )^2 .
\label{trans}
\end{equation}
This model too has an infrared massless fixed point but this is now dominated by a quartic term, not a quadratic one. Note that this is an allowed world-sheet theory of a string model since translation invariance in embedding space is now respected after identifying $\chi$ with the embedding coordinates.

We now show that the massive extension of the Polyakov action and the Lifshitz model are dual to each other. To this end we consider the master Lagrangian
\be
{\cal L}=\frac{1} {2} v^\mu v_\mu -{\rm i} \varphi \epsilon^{\mu\nu}\partial_\mu v_\nu+m^2 \varphi^2 \ ,
\ee
where $v_\mu$ is a vector field. For $m=0$ we get the standard $T$-duality: integrating out $v_\mu$ we get the free action for $\varphi$ but integrating out $\varphi$ we  get $\epsilon^{\mu\nu}\partial_\mu v_\nu=0$ which locally implies $v_\mu = \partial_\mu \chi$ giving rise to a free action for the dual scalar field $\chi$. We then get the two dual free theories for which the Lagrangians and operators are related as follows:
\be
{\cal L}(\varphi)=\frac{1}{ 2}\partial^\mu\varphi \partial_\mu \varphi\Longleftrightarrow {\tilde{\cal L}}(\chi) =\frac{1}{ 2}\partial^\mu\chi \partial_\mu \chi \ ,\qquad \partial_\mu\varphi \Longleftrightarrow \epsilon_{\mu\nu}\partial^\nu \chi \ .
\ee

For the massive case $m\neq 0$, by the same procedure, we obtain the duality
\bea
{\cal L}(\varphi)=\frac{1}{ 2}\partial^\mu\varphi \partial_\mu \varphi+m^2\varphi^2 &\Longleftrightarrow & {\tilde{\cal L}}(\chi)=\frac{1}{ 2}\partial^\mu\chi \partial_\mu \chi+\frac{1}{2m^2}\left(\nabla^2\chi\right)^2,\nonumber \\
\partial_\mu\varphi  &\Longleftrightarrow & \epsilon_{\mu\nu}\partial^\nu \chi \ ,
\label{dualmassive}
\eea
relating a standard massive scalar field theory to a Lifshitz high derivative theory after identifying $v_\mu=\partial_\mu \chi$ since in two dimensions a massive vector carries only one degree of freedom.\footnote{For a recent discussion  of general duality transformations among antisymmetric tensor field theories mapping massive theories to higher derivative theories see \cite{Burgess:2025geh}.} In the latter formulation, the point $m^2=0$ is dominated by a quartic derivative interaction.

Unfortunately, when the (Euclidean) time and space derivatives enter symmetrically, the Lifshitz model cannot be Wick rotated to Minkowksi space-time without the appearance of ghosts, states of negative norm, and/or tachyons. Essentially, the Lifshitz model represents the original rigid string of Polyakov and Kleinert \cite{Polyakov:1986cs,Kleinert:1986bk}, which is known to be plagued by ghosts.

Fortunately, there is a way to fix the problem as proposed in \cite{Diamantini:1998zu} and further studied in \cite{Diamantini:1999uj}. The idea is simply to add one extra higher derivative term to the Lifshitz model. This term may also permit a negative sign for the quadratic term, inducing a roton in the Euclidean spectrum. This is the situation considered originally by Brazovskii \cite{brazovskii1975phase}, and leading to
\begin{equation}
{\cal L} =  \partial_{\mu} \chi \left( T - S\nabla^2 +\frac{1} {M^2} {\nabla}^4 \right) \partial_{\mu}\chi \ .
\label{sixth}
\end{equation}
In terms of the original $\chi$ fields, this model involves an extra  sixth-order term in derivatives. It is exactly this term that fixes the ghost problem (even though this modification seems completely unjustified now, we will see later in section \ref{sec3} that the terms within the round brackets can be interpreted as the first few terms in the expansion of a non-local kernel on a worldsheet surface). If the condition (this condition will follow from the non-local kernel derived in section \ref{sec3} as well)
\begin{equation}
T -Sq^2 +\frac{1} {M^2}q^4 > 0, \qquad \forall q^2 \in {\mathbb R} \ ,
\label{ghost}
\end{equation}
is satisfied, the model has a single pole at $q^2 = 0$ and no tachyons nor ghosts appear upon a Wick-rotation to Minkowski space time. For $S<0$, as originally considered by Brazovskii \cite{brazovskii1975phase},
the Euclidean spectrum develops a roton minimum at finite momentum $q^2 \approx SM^2/3$, indicating an instability toward formation of microstructure on this scale.

When the fields $\chi$ are interpreted as the embedding fields $x^{\mu}$ of a string theory, the Brazovskii-Lifshitz model becomes exactly the string theory with hyperfine structure introduced in \cite{Diamantini:1999uj}.

Based on these observations, in the next subsection we will review the string action introduced in \cite{Diamantini:1998zu} and further studied in \cite{Diamantini:1999uj}.

\subsection{Brazovskii-Lifshitz strings}

The string action proposed and studied in \cite{Diamantini:1998zu, Diamantini:1999uj} to describe the flux tubes of the confining phase of gauge theories is
\begin{equation}
S = \int d^2{\sigma} \sqrt{g} \ g^{ab}
{\cal D}_a x_{\mu }\! \left( T - S\,{\cal D}^2 +
\frac{1} {M^2} {\cal D}^4 \right)
\ {\cal D}_b x^{\mu } \ ,
\label{one}
\end{equation}
where ${\cal D}_a$ are covariant derivatives with respect to the induced metric $g_{ab}=\partial_a x_{\mu }\partial _b x^{\mu }$ on the surface $x^{\mu}\left( \sigma^0, \sigma^1\right) $ embedded in $D=d+1$ dimensions and ${\cal D}^2$ the corresponding covariant Laplacian
\begin{equation}
{\cal D}^2 = \frac{1}{ \sqrt{g}} \partial _a g^{ab} \sqrt{g}\partial _b \ .
\label{covlap}
\end{equation}
The first term provides a bare surface tension $2T$, the second accounts for rigidity with stiffness parameter $S$, called the string fine structure \cite{Polyakov:1986cs,
Kleinert:1986bk} and, finally, the last term is the string hyperfine structure, defined by the mass scale $M$. Without the quartic derivative term within the round brackets, \eqref{covlap} is the rigid string originally introduced in \cite{Polyakov:1986cs, Kleinert:1986bk}, see also \cite{Kleinert:1987}. In this case, the stiffness can only be positive. The presence of the hyperfine structure allows also a negative stiffness, which seems to be a crucial feature of confining strings \cite{Kleinert:1996, Diamantini:1997wk}.

The hyperfine structure eliminates also the ghosts plaguing the rigid string.
If the condition (\ref{ghost}) is satisfied the model has a single pole at $q^2 = 0$ and no tachyons nor ghosts  appear upon a Wick-rotation to Minkowski space time, as we show in detail below. This condition is automatically satisfied in the physical range of the model. Note that, due to the hyperfine structure, one can admit also negative stiffness $S<0$. The Euclidean roton at finite momentum $q^2 \approx SM^2/3$, leads to a modulated ground state of the world sheet, akin to an ``egg-carton", which is the original reason for the name ``hyperfine structure" \cite{Diamantini:1998zu}.

The quartic hyperfine structure appears as perturbatively irrelevant and, as such, it is typically dismissed. Actually, it turns out that this is a dangerously irrelevant operator that dominates the entire string theory in the infrared in the saddle-point approximation \cite{Diamantini:1998zu}. One can show, by integrating over fluctuations, that the hyperfine term induces a dynamical string tension $\propto m^2 \Lambda^2$, where $m$ is a dimensionless parameter and $\Lambda$ the necessary cutoff. The cutoff can then be removed and the parameters flow to an infrared fixed point $t=T/\Lambda^2 \to 0$, $S\to 0$ and $m\to 0$ with $t/m^2 \to 0$ as we will illustrate below in section \ref{sec4}. At this fixed point the string can be renormalized, so that a mass scale $M_{\rm ren}$ emerges from the simultaneous limit $\Lambda \to \infty$ and $m\to 0$. The hyperfine term generates a dynamical string tension $T_{\rm ren}\propto M_{\rm ren}$ that takes over the oriental correlations where the bare Nambu-Goto tension fails to prevent crumpling, so that the string world-surface has Hausdorff dimension $d_{\rm H}=2$, as it should be for a well-defined flux tube. The Euclidean propagator takes the form $\propto 1/p^2(M_{\rm ren}^4 + p^4)$. Correspondingly, the standard deviation of transverse displacements is automatically regulated so that it takes the form $\sigma_{\rm trans} \propto {\rm log}\  M_{\rm ren}L$, showing that $1/M_{\rm ren}$ is the intrinsic thickness of the string. This is an infrared Brazovskii-Lifshitz fixed point dominated by the highest-order hyperfine term.
For strings much longer than $1/{M_{\rm ren}}$ one obtains back Nambu-Goto behaviour with a central charge $c=1$ per transverse degree of freedom. For shorter strings, instead, there are sizeable deviations.

Note that this behaviour does not depend on the particular truncation of a higher derivative expansion to a quartic potential. The same results are obtained for all consistent truncations and lead to the same ghost- and tachyon-free fixed point\cite{Diamantini:1999uj}.

Finally, let us stress a  crucial difference to the Polyakov string. Even if we introduce a Lagrange multiplier for an independent world-sheet metric \`a la Polyakov, here this metric survives also inside the powers of the Laplacian in the higher-order terms. Therefore there is no Weyl symmetry, even at the classical level. The only symmetries are reparametrization invariances on the world sheet, which can be completely fixed in transverse gauge by choosing one space coordinate along the string and writing an action for the remaining transverse coordinates alone. Therefore, there is also no need to cancel a Weyl anomaly at quantum level which implies the absence of any constraint on the allowed space-time dimensions: 3+1 dimensions are perfectly fine. The absence of Weyl symmetry has another important consequence. The Virasoro constraints of vanishing energy-momentum tensor are also absent. Therefore, also the usual mass formula following from these constraints is absent and, correspondingly there is no tachyon in the target space spectrum. Both the Polyakov string and the Brazovskii-Lifshitz string introduced in \cite{Diamantini:1998zu} are quantum consistent strings\footnote{This statement refers to the absence of tachyons and ghosts in the free spectrum and, as shown in section \ref{sec5}, at leading order in the $1/D$ saddle-point expansion. Since the physically relevant case has $D-2=2$ transverse dimensions, i.e.\ $D=4$, which is not parametrically large, an independent demonstration of unitarity beyond this leading order is not attempted here; see the footnote at the start of section \ref{sec5} for further discussion.}. The Polyakov string is stabilized by a critical dimension 26, the Brazovskii-Lifshitz string by a finite thickness.

Summarizing, the quartic hyperfine term achieves:
\begin{itemize}
    \item{An IR string fixed point describing a zero stiffness, $c=1$ conformal field theory for long strings with fixed deviations appearing on shorter scales.}
    \item{The geometry is that of a world-sheet with Hausdorff dimension $d_{\rm H}=2$ and an intrinsic thickness.}
    \item{The action can be enhanced by adding infinite higher order terms. Each consistent truncation leads to the same ghost- and tachyon-free non-perturbative fixed point.}
\end{itemize}

Note however that these interesting properties come from adding an extra six-derivative term to the action without a derivation for the higher derivative terms. In the next section, we will consider compact QED including a topological $\theta-$term in the phase in which dyonic monopoles condense and we will show that this model reproduces exactly to the BL string studied above.

\section{Compact QED with \texorpdfstring{$\theta$}{theta}--Term and Confining Strings}\label{sec3}


In this section we review $U(1)$ compact QED in the presence of a $\theta-$term. We briefly discuss how, even in the presence of this extra term, the remarkable strong-weak-coupling electromagnetic duality does persist in the form of a $SL(2,\mathbb{Z})$ group. We also review the dual string formulation of compact QED in the monopole condensation phase \cite{Quevedo:1996uu}.

\subsection{Compact QED and Duality}

A remarkable feature of compact Maxwell electromagnetism in the presence of magnetic charges is its strong-weak duality under interchange of the electric and magnetic fields, as epitomized by the Dirac quantization condition. Whether or not these magnetic monopoles exist remains unknown, since direct experimental evidence is still lacking. In addition to monopoles, there is another unique feature, the fact that this compact $U(1)$ theory allows a topological term characterized by a $\theta$ parameter. It turns out that this extra term does not spoil the duality just mentioned. Actually, this is enhanced to invariance under $SL(2,\mathbb{Z})$ transformations of a complex parameter $\tau = \theta/2\pi + 2 \pi i/e^2$. This can be seen by considering the action of compact $U(1)$ gauge theory in 4d Euclidean space including the topological term,
\begin{equation}\label{cQED}
S = \int d^4x \left( \frac{1}{4 e^2} F_{\mu \nu} F^{\mu \nu} + \frac{i \theta}{16 \pi^2} F_{\mu \nu} \tilde{F}^{\mu \nu} \right)\,,
\end{equation}
where the normalization of the $\theta$-term has been chosen such that under a shift $\theta \to \theta +2 \pi$ the physics does not change. This is because the second Chern class $c_2(F) = (1/8\pi^2) F\wedge F$ gives always an integer when integrated over a manifold. The elementary shift in the $\theta$-parameter becomes a shift in the complex parameter, $\tau \to \tau +1$. To fully exhibit the $SL(2,\mathbb{Z})$ invariance, we have to show that there is a hidden symmetry under $\tau \to -1/\tau$ in equation \eqref{cQED}. A convenient way to show this is by introducing a basis of dual and anti-self dual forms as $F^{\pm} = (F \pm \ast F)/2$ and writing
\begin{equation}
S(F,\tau) = \frac{i} {4\pi} \int ( \bar{\tau} (F^+,F^+) -  \tau (F^-,F^-)) \ ,
\label{elac}
\end{equation}
where we have used the definition of the inner product of differential forms. We now introduce a generic two-form $B$, a new, ``magnetic" $U(1)$ connection with associated curvature two-form $G$ and we consider the master action
\begin{equation}
S_{\rm master} = S(B, \tau) + \frac{i} {2\pi} \int B \wedge G \ .
\label{master}
\end{equation}
By integrating over the new connection we obtain the constraint that $B$ must be an exact form $F$ and we get back the original ``electric" action (\ref{elac}). On the other side we can also eliminate $B$ in favour of $G$ by a simple Gaussian integration. This gives the ``magnetic" action
\begin{equation}
S(G, \tau) =\frac{i} {4\pi}  \int \left(-\frac{1} {\bar{\tau}} (G^+,G^+) + \frac{1} {\tau} (G^-,G^-)\right) \ ,
\label{maac}
\end{equation}
i.e. $S(G,-{1/\tau})$. This is the same action as the original (\ref{elac}) but with coupling $-1/\tau$, which proves the full $SL(2,{\mathbb Z})$ duality.

Dualities like these have played a fundamental role in physics and, as demonstrated by section \ref{sec2}, are also playing a key role in this work. Now we will review how compact QED in the monopole condensation phase is related to a confining string theory.

\subsection{Monopole Condensation and Confining Strings}

In the presence of monopoles, the $\theta-$term leads to non-trivial effects rather than just being a boundary term. For example, monopoles acquire an electric charge $e \,\theta/2\pi$ \cite{Witten:1979ey}. Moreover, it turns out that, at low energies, the effects of a monopole condensation can be captured by promoting the exact form $F$ to a fundamental field $B$ \cite{Quevedo:1996uu}\footnote{This expression can actually  be derived by dualizing compact QED and considering monopole condensation as in Polyakov's original work on confinement (see for instance \cite{Polyakov:1996nc,Quevedo:1996uu}). The structure of massive 2-forms generalizes the massive scalar field dualities in 2d as discussed in section 2.}. Thus, instead of considering \eqref{cQED} we focus on the gauge theory for a $2-$form field
\begin{equation}\label{cQEDplusH}
S[B] = \int d^4x \left( \frac{1}{12 z^2 \Lambda^2} H_{\mu \nu \rho}H^{\mu \nu \rho}\left(1+{\mathcal{O}}\left(\frac{H^2}{\Lambda^2}\right)\right)
+\frac{1}{4 e^2} B_{\mu \nu} B^{\mu \nu} + \frac{i \theta}{16 \pi^2} B_{\mu \nu} \tilde{B}^{\mu \nu} \right)\,,
\end{equation}
with $H_{\mu \nu \rho} = \partial_{\left[ \mu \right.} B_{\left. \nu \rho \right]}$ an antisymmetric $3-$form, $\Lambda$ a new coupling constant with dimension of mass, associated to the density of the condensation and $z\propto {\rm exp} (-{\rm const}/e^2)$ the monopole fugacity. The $e$-dependence of the fugacity $z$ makes it explicit that  monopole condensation is a non-perturbative effect. Of course, the full action obtained by duality involves a specific infinite sum of powers of $H_{\mu \nu \alpha}H^{\mu \nu \alpha}/\Lambda^2$ \cite{Diamantini:1996vf}. Here we retain only the low-energy Gaussian approximation since we are interested in the limit $\Lambda\rightarrow\infty$. We will see below that this Gaussian duality becomes exact in presence of a $\theta-$term, since it becomes possible to remove the cutoff.

Noting that only transverse modes are physical, we can introduce a vector field $a^{\mu}$ by $H_{\mu \nu \rho} = \epsilon_{\mu \nu \rho \sigma} a^{\sigma}$ and show that there is one massive vector degree of freedom whose equation of motion is $(\partial_{\nu} \partial^{\nu} + m_{\theta}^2) a^{\mu} = 0$, with mass $m_{\theta}$ given by
\begin{equation}
m_{\theta} = \frac{e z \Lambda}{2 \pi} \sqrt{\left(\frac{2 \pi }{e^2} \right)^2 + \left(\frac{\theta }{2 \pi} \right)^2} = \frac{ez\Lambda}{2\pi} |\tau|\ .
\end{equation}
The additional degree of freedom with respect to the original massless photon represents the fluctuations of the monopole condensate \cite{Quevedo:1996uu}. In the absence of the $\theta-$term, this would correspond to a massive $B-$field with mass $m = z\Lambda/e$ inversely proportional to the coupling and therefore vanishing in the strong coupling limit. The presence of a $\theta$-angle reverses this relation; for strong coupling the mass becomes $m_{\theta} \approx ez\theta \Lambda/4\pi^2$ which is linearly divergent in the coupling. This happens for any non-vanishing value of $\theta$ although only $\theta = \pi$ preserves parity and time-reversal invariance. This observation plays a fundamental role in this work, so we will return to this later.

Since an antisymmetric two-index tensor $B_{\mu\nu}$ naturally couples to a two-dimensional extended object, it turns out that this model is naturally a model of strings \cite{Diamantini:1996vf}, so that we should include the coupling to a string source
 \begin{equation}\label{source}
 J^{\mu \nu} = \frac{1}{2} \int d^2\sigma\, T^{\mu \nu}\, \delta^{(4)}({x - x(\sigma))}\,, \qquad T_{\mu \nu} = \epsilon^{ab} \partial_a x_{\mu} \partial_b x_{\nu}\ ,
 \end{equation}
we then consider the following action rather than \eqref{cQEDplusH}
\begin{eqnarray}
    S[B,J] = S[B] - i \int d^{4}x \ B_{\mu \nu} J^{\mu \nu}\,.
\end{eqnarray}
From the previous discussion we emphasize that the $\theta-$term turns the massive $B-$field into a heavy field at strong gauge coupling -- consistent with the non-perturbative phenomenon of monopole condensation -- allowing us to integrate it out and finding an effective action for the string sources. The effective Euclidean action of the string in $\exp(-S_{\text{eff}}[J])$ resulting from integrating out the $B-$field, since this is a simple Gaussian integration, is given by \footnote{Note that the current $J$ coupling to the $2$-form field $B$ describes no more particles, but rather strings sweeping a world-sheet parameterized by ${\bf{x}} (\sigma)$. Importantly, this current $J$ is not added in by hand but it emerges upon duality as a representation of the sum over periods of the original cosine function of compact QED \cite{Diamantini:1996vf}.}
\begin{equation}\label{effac}
    S_{\text{eff}}[J] = \frac{z^2 \Lambda^2}{2}\int d^4x\, d^4y \,J^{\mu \nu}(x)\,\mathsf{K}(x,x)J_{\mu \nu}(y)\,,
\end{equation}
where $\mathsf{K}(x,x)$ is a non-local kernel in target spacetime that can be obtained by inverting the Fourier transform of
\begin{equation}
    \mathsf{K}(x,y) =
\int \frac{d^4k}{(2 \pi)^4} \frac{e^{ik\cdot(x-y)}}{{k^2 +m_{\theta}^2}}\,.
\end{equation}
This corresponds to the $4-$dimensional Yukawa Green's function that can always be expressed in terms of a modified Bessel function. Plugging \eqref{source} in equation \eqref{effac} leads to a kernel $\mathsf{K}(x(\sigma),y(\sigma))$. We would like to express this as a non-local kernel on the string world-sheet because we are interested in deriving the action of the \emph{confining} string. For this purpose we shall introduce coordinates $\sigma^0, \sigma^1, \chi^2,  \chi^3$ so that $\chi^i$ describe the 2 normal directions to the world-sheet. Using these coordinates, the kernel can be expressed as \cite{Diamantini:1997wk}
\begin{align}\label{kernelS}
    \mathsf{K}(x(\sigma),x(\sigma')) = \int d^{2}\chi \, \delta^{(2)}(\chi) \,\frac{1}{m_{\theta}^2 - \nabla^2_{\chi}} \, \delta^{(2)}(\chi)\, \frac{1}{\sqrt{g}}\, \delta^{(2)}(\sigma - \sigma')\,,
\end{align}
with $g={\rm det}\ \partial_a x_{\mu}\partial_b x^{\mu}$ the determinant of the induced metric and
where we have used $(-\nabla^2 + m_{\theta}^2){\sf K}(x,y) = \delta^{4}(x-y)$ and the fact that we can express the target spacetime delta function as a product of delta functions for tangent and transverse coordinates up to the Jacobian associated with this transformation \cite{Diamantini:1997wk}. This expression is plagued by an ultraviolet divergence that has to be properly regularized. With an appropriate choice of regularization, this leads to a well-behaved Green's function \cite{Diamantini:1997wk}
\begin{align}\label{greenS}
    \mathsf{K}(x(\sigma),x(\sigma')) = G\left(\rho(e),\left(\frac{\mathcal{D}}{\Lambda}\right)^2\right)\, \frac{1}{\sqrt{g}} \, \delta^{(2)}(\sigma - \sigma') \,,
\end{align}
where $\mathcal{D}^2= \partial_a(\sqrt{g}\,g^{ab}\partial_b)$ is the surface covariant derivative, $\Lambda$ is the cutoff scale of the original gauge theory and $G$ is defined as the Taylor series obtained from
the generating function
\begin{equation}
G\left(z,\rho(e),
\left( \frac{{\cal D}}{ m_{\theta}}\right) ^2\right)=
\frac{z^2}{{4 \pi}}\ K_0 \left( \rho(e) \sqrt{1
- \left( \frac{{\cal D}}{m_{\theta}}
\right)^2 } \right) \ ,
\label{green}
\end{equation}
with
\begin{equation}
\rho (e) = \frac{m_{\theta}}{\Lambda} \ ,
\label{tau}
\end{equation}
the ratio of the string mass gap $m_\theta$ and the gauge cutoff and  $K_0$ the zeroth-order modified Bessel function of the second kind.

Inserting \eqref{source}, \eqref{greenS} into \eqref{effac} one obtains the confining string action
\begin{equation}\label{csa}
    S_{\text{cs}} = \Lambda^2 \int d^2\sigma \,\, \sqrt{g}\, \, t_{\mu \nu} (\sigma)\, G\left(\rho(e),\left(\frac{\mathcal{D}}{m_{\theta}} \right)^2 \right) \, t^{\mu \nu}(\sigma)\,.
\end{equation}
where
\begin{equation}
t_{\mu \nu} = \frac{1} {\sqrt{2g}} \,T_{\mu \nu} \ ,
\label{tang}
\end{equation}
is the normalized tangent tensor to the world-sheet and $G$ has to be understood as a generating function for its derivative expansion.

When a $\theta-$term is present, there is an additional contribution to the world-sheet action, given by
\begin{equation}
\Delta S_{\theta}= i \frac{\theta}{ 4\pi \left( \left( \frac{2\pi} {e^2} \right)^2 + \left(\frac{\theta}{2\pi}\right)^2 \right)} \nu \ ,
\label{addtheta}
\end{equation}
where 
\begin{equation}
\nu=\frac{1}{4\pi} \int d^2 \sigma \sqrt{g} \ \epsilon^{\mu \nu \alpha \beta} g^{ab} \partial_a t_{\mu \nu} \ \partial_b t_{\alpha \beta} \ ,
\label{sin}
\end{equation}
is the (signed) self-intersection number of the world-sheet in 4d Euclidean spacetime \cite{Polyakov:1987hqn}. In the limit $e^2 \to \infty$ and for $\theta = \pi$ the world-sheets acquire a minus sign for each self-intersection, realizing what Polyakov has called ``fermionic strings".

Finally, to make contact with the Brazovskii-Lifshitz string discussed in the previous section, we would like to rewrite this action in terms of tangent vectors ${\cal D}_a x_{\mu}$ to the world-sheet. To this end we consider
a generic term in the expansion of the kernel (\ref{green}) in terms of ${\cal D}^2$:
\begin{equation}
{t}_{\mu \nu } {\cal D}^{2k} t_{\mu \nu } = \frac{1}{2}\left(
g^{ac} g^{bd}- g^{ad} g^{bc} \right) \partial_a x_{\mu }\partial_b x_{\nu }
\left( {\cal D}^{2k} \partial _cx_{\mu } \partial _d x_{\nu } +
\partial _c x_{\mu } {\cal D}^{2k}\partial _d x_{\nu } + r_{k;cd\mu \nu }
\right) \ .
\label{exp1}
\end{equation}
Here $r_{k;cd\mu \nu }$ represents the additional terms where the $2k$
covariant derivatives are distributed among $\partial _c x_{\mu }$ and
$\partial _d x_{\nu }$. Using the definition of induced metric we obtain
\begin{equation}
{t}_{\mu \nu } \ {\cal D}^{2k}\ t_{\mu \nu } = g^{ab} \ {\cal D}_a
x_{\mu }{\cal D}^{2k} {\cal D}_b x_{\mu } + r_k \ ,
\label{exp2}
\end{equation}
where $r_0=-1$ and $r_1=0$. At fixed $k$, the first, ``diagonal" terms contain the minimal number $(2k+2)$ of world-sheet derivatives, while the remainders necessarily involve additional derivatives beyond this minimal structure since there are fewer metric contractions. Since we are interested only in large-distance properties we shall thus neglect the sub-dominant remainders for $k\ge 2$. Note, however, that the power-counting irrelevant diagonal terms may well become relevant non-perturbatively, when the sum over transverse directions can compensate the derivative-induced suppression.  This gives
\begin{equation}\label{csader}
    S_{\text{cs}} = \Lambda^2 \int d^2\sigma \,\, \sqrt{g}\, \, g^{ab}{\cal D}_a x_{\mu} V\left(z,\rho(e),\left(\frac{\mathcal{D}}{m_{\theta}} \right)^2 \right) \, {\cal D}_bx^{\mu} (\sigma)\,.
\end{equation}
where,
\begin{equation}
V\left(z,\rho(e),
\left( \frac{{\cal D}} {m_{\theta}}\right) ^2\right) = \left(
G\left(z,\rho(e),
\left( \frac{{\cal D}}{ m_{\theta}}\right) ^2\right)-\frac{1}{2}
G\left(z,\rho(e),0\right) \right) \ .
\label{finalgreen}
\end{equation}
We shall further study this action in the next section, describing its connection to the Brazovskii-Lifshitz string. In doing so we omit the topological term which plays no role for the bulk string dynamics.

We would like to stress that this kernel represents the exact world-sheet action in the Gaussian approximation for the two-form field generated by monopole condensation. As such, it generates higher-derivative terms but not higher-power interactions for fields. One could loosely call it a ``non-local Gaussian approximation" for the world-sheet of the confining string action.

\section{Brazovskii-Lifshitz Strings from Compact QED with a \texorpdfstring{$\theta$}{theta}--Term}\label{sec4}

In this section we will establish the connection between the confining string action obtained in the previous section through compact $U(1)$ and the BL string introduced in section \ref{sec2}.

In \cite{Quevedo:1996uu,Polyakov:1996nc,Diamantini:1996vf}, and as reviewed in the previous section, it was shown that compact QED in the monopole condensation phase has an exact dual formulation as a closed string theory, in which a possible $\theta$-term is matched to a topological term measuring the self-intersection number of the world sheet.
Here we shall show how this leads exactly to the previously discussed Brazovskii-Lifshitz string.

The action \eqref{csader} can be rewritten (up to the topological term, which we will always consider as added in) as
\begin{eqnarray}
S_{\rm cs} &&= \int d^2{\sigma} \sqrt{g} \ g^{ab} {\cal D}_a x_{\mu }\ \left( T - S\,{\cal D}^2 +\frac{\Delta }{ m_{\theta}^2}\, {\cal D}^4 \right) \ {\cal D}_b x_{\mu } \ ,
\nonumber \\
\Delta &&= c_4+ c_6 \frac{1}{ m_{\theta}^2} {\cal D}^2 + c_8 \frac{1}{ m_{\theta}^4} {\cal D}^4 + \dots \ ,
\label{exp}
\end{eqnarray}
where the first few coefficients are given by
\begin{eqnarray}
T &&=\frac{z^2}{ {8 \pi }} \Lambda^2 \ K_0 (\rho ) \ ,
\nonumber \\
S &&=-\frac{z^2}{ {16 \pi }}\ \frac{1}{ \rho} \ K_1 (\rho ) \ .
\nonumber \\
c_4 &&=\frac{z^2}{ {32 \pi }} \ K_2 (\rho ) \ ,
\label{coeff}
\end{eqnarray}

Note that since the non-local action \eqref{csader} was obtained by a proper Gaussian integration of a heavy field, the higher derivative theory is automatically free of ghosts and tachyons. Recall that for the field $B_{\mu\nu}$ to be heavy at strong coupling $e$ we need $\theta\neq 0$.

Consider now the series $\Delta$ in momentum space, where it becomes a function of $q^2/m_{\theta}^2$. Since it sums up to the generating function (\ref{finalgreen}) involving the Bessel function (\ref{green}), momenta must be restricted to avoid divergences. So,
let us assume that we integrate over momenta up to $|q|_{\rm max}$. In this way we recognize that the model involves a second dimensionless parameter $\kappa = q_{\rm max}^2/m_\theta^2$, in addition to the coupling constant $e^2$ contained in $\rho(e)$. This new parameter measures the scale on which one probes the ultraviolet physics.

We now re-package the non-perturbative effects of the whole infinite series $\Delta$ by promoting it to a parameter $\Delta(\rho (e), \kappa)$. As a consequence we obtain a Brazovskii-Lifshitz string (\ref{one}) with $M^2=m_\theta^2/\Delta$ and coefficients depending on two parameters, $e^2$, the gauge coupling, and $\kappa$. Note that the stiffness $S$ is negative and thus this string has hyperfine structure.

The important points are two.
\begin{itemize}
\item First of all, note that the full kernel (\ref{green}) (and thus also (\ref{finalgreen})) diverges logarithmically for $\kappa \to 1_{-}$. Since the bare tension $T$ and stiffness $S$ are finite for $\kappa \to 1_{-}$, this means that also the coefficient $\Delta(\rho, \kappa)$ diverges logarithmically for $\kappa \to 1_{-}$.
\item Secondly, for $\theta=0$ the mass ratio $\rho =z/e$ vanishes for strong coupling, so that $K_0(\rho)$ diverges in this limit. As a consequence, the bare tension diverges too and cannot be renormalized.
For non-zero $\theta$, however the mass ratio behaves as
\begin{equation}
\rho(e)  \approx \frac{ez\theta}{ 4\pi^2} \ ,
\label{thetamass}
\end{equation}
for $e\gg 1$.  In this case, $\rho$ diverges in the strong coupling limit and the bare tension can be renormalized when we remove the cutoff because $K_0$ vanishes as $K_0(\rho) \approx {\rm exp}({-\rho})/\sqrt{\rho}$ for large values of $\rho$. Since this is also the large-value asymptotics of $K_1$, the stiffness $S$ vanishes for large coupling.
\end{itemize}

Care is needed in treating the quartic term, though.
All the coefficients $c_{2n}$ involve integer Bessel function up to $K_n$, with large-value asymptotics the same as for $K_0$, so that the factor $\Delta(\rho, \kappa)$ vanishes in the limit $\rho \to \infty$ at fixed $\kappa $. However, in the limit $\kappa \to 1_{-}$ at fixed $\rho$, $\Delta$ diverges logarithmically, as we have pointed out above. The limit we are interested in is thus, $\rho\to \infty$ and $\kappa \to 1_{-}$ simultaneously so that $m_\theta^2 / \Delta = \Lambda^2 \rho^2 /\Delta \to {\rm finite}\ M_{\rm ren}^2$ as the cutoff is removed, $\Lambda \to \infty$, which requires $\rho^2/\Delta$ to vanish so that we obtain a finite, renormalized mass  $M_{\rm ren}^2$.
Since
\begin{equation}
\frac{T}{ M_{\rm ren}^2} = \frac{z^2}{ 8\pi} \frac{\Delta\left(\rho,\kappa \right) K_0\left(\rho\right)}{ \rho^2 } \ ,
\label{massratio}
\end{equation}
requiring $\Delta(\rho,\kappa)/\rho^2 \to \infty$ but $\Delta(\rho,\kappa) K_0(\rho) /\rho^2 \to 0$ for $\rho \to \infty$ and $\kappa \to 1_{-}$ we obtain exactly the Brazowskii-Lifshitz string of the previous section in which the renormalized bare tension $T_{\rm ren}$ vanishes on the scale of the renormalized mass $M_{\rm ren}^2$ and the string tension in the continuum is entirely due to the quartic interaction. Note that the power-counting irrelevant quartic operator becomes non-perturbatively relevant due to the divergence of the infinite sum encoding the string physics up to the momentum cutoff. This describes an IR strong coupling ($e\to \infty$) fixed point dual to compact QED. In the next section we shall present arguments to suggest that this is a universal confinement fixed point in which the only remnant of the original colour group is encoded in the renormalized mass $M_{\rm ren}$ which describes a finite string width \cite{Diamantini:1998zu}.

Summarizing: At strong coupling $e\rightarrow \infty$ and nonvanishing $\theta$ we have that the fugacity $z\sim e^{-1/e^2}\rightarrow 1$ but the mass $m_\theta\sim \frac{\Lambda z\theta e}{4\pi^2}\rightarrow \infty$ and also the ratio $\rho=\frac{m_\theta}{\Lambda}\rightarrow \infty$. Recalling that we can consider an independent limit  $\kappa=\frac{q^2_{\rm max}}{m_\theta^2}\rightarrow 1_-$  we can play  with a double scaling  limit in order to eliminate the dependence on the cut-off scale $\Lambda$ and have a finite value of the renormalized mass scale $M_{\rm ren}^2=\frac{m_\theta^2}{\Delta}$. With the couplings $S,T,\Delta$ in the high derivative expansion and the renormalized  mass $M_{\rm ren}$ behaving as in the following table.

\begin{equation}
\begin{array}{|c|c|c|c| c|}
 \hline &\rho\, {\rm fixed}, \kappa\rightarrow 1_-& \rho\rightarrow \infty, \kappa<1& \rho\rightarrow \infty, \kappa \rightarrow 1_- \\ \hline
 T& {\rm finite} & 0& 0\\
S& {\rm finite} & 0& 0\\
\Delta &\infty & 0&\infty \\
M_{\rm ren} & 0 &\infty & {\rm finite}\\ \hline
\end{array}
\end{equation}
The continuum limit above realizes dimensional transmutation. For nonzero
$\theta$, sending $e^2 \to \infty$ and $\Lambda \to \infty$ gives
$\rho(e) \to \infty$ and $m_\theta = \Lambda \rho(e) \to \infty$. Since
$q_{\max} = \sqrt{\kappa}\,m_\theta$, the simultaneous limit
$\kappa \to 1_{-}$ also sends $q_{\max} \to \infty$: both dimensionful
cutoffs are removed. Their correlated scaling is determined by retaining
a finite physical mass,
\begin{equation}
    M^2_{\rm ren}
    = \lim_{\Lambda \to \infty}
    \frac{\Lambda^2 \rho^2}{
    \Delta\bigl(\rho,\kappa\bigr)},
    \qquad
    0 < M^2 < \infty,
\end{equation}
with $T/M^2_{\rm ren} \to 0$ and $S \to 0$ as specified above. Thus the
dimensionless microscopic coupling is traded for a dimensionful scale,
in the same sense as dimensional transmutation in Yang--Mills theory,
although the present limit is at strong bare coupling. The surviving
scale is $M_{\rm ren}$, not the divergent bare mass $m_\theta$; it sets the
intrinsic string width, proportional to $(M_{\rm ren})^{-1}$, and the dynamically
generated tension, proportional to $M^2_{\rm ren}$.

The condition $\kappa \to 1_{-}$ expresses the scaling required to retain
this finite physical scale and does not leave a finite momentum cutoff.
The resulting Brazovskii--Lifshitz fixed point identifies an entire
universality class of microscopic compact-QED realizations whose induced
string actions approach $T/M^2_{\rm ren} = S = 0$ and differ only by irrelevant
terms, at fixed topological sector. The persistence of this fixed point
across consistent higher-derivative truncations, reviewed in
Section~\ref{sec2} \cite{Diamantini:1998zu,Diamantini:1999uj}, supports this independence from microscopic details.
Universality concerns this critical domain; it does not require attraction
from arbitrary microscopic couplings without tuning. Hence the correlated
scaling is the renormalization prescription selecting the continuum
universality class, with $M_{\rm ren}$ setting its physical units. In what follows,
we denote the renormalized mass simply by $M$.

\section{Saddle-point Analysis of the Confining String}\label{sec5}

In this section, we shall compute the effective action including the one-loop contribution coming from integrating out the transverse modes of an open string of finite length in $D-$dimensional spacetime. To do so we will follow the classic non-perturbative quantization of strings introduced in \cite{Alvarez:1981kc}.\footnote{The $1/D$ expansion used throughout this section is a systematic saddle-point scheme that is parametrically controlled for $D\gg1$. The physically relevant case, $D=4$ (i.e.\ $D-2=2$ transverse fluctuating coordinates), is not parametrically large, so the results below should be understood as the leading-order term of this expansion, in line with the rest of the effective-string literature that uses the same scheme; we do not attempt an independent estimate of the size of subleading $1/D$ corrections at $D=4$.} This requires enforcing the induced metric constraint by a Lagrange multiplier, integrating out the $(D-2)$ transverse modes and deriving the $1/D$ expansion for the remaining fields. We shall discuss the string spectrum in the fixed point and make some comments regarding proposals in the literature. Moreover we will focus, in particular on the saddle-point gap equations representing the dominant term in the $1/D$ expansion and we will solve them numerically. We will mostly follow \cite{Diamantini:1999uj} but extend their saddle-point analysis to include the full one-loop contribution and not only the dominant IR term.

\subsection{String spectrum in the fixed point}\label{spectrums}

Let us now derive the particle content of the Brazovskii-Lifshitz string. A major problem of the rigid string \cite{Polyakov:1986cs,
Kleinert:1986bk} is the presence of an unphysical world-sheet ghost in Minkowski space-time \cite{Kleinert:1986bk}. One of the motivations for introducing the hyperfine string structure \cite{Diamantini:1998zu} was to replace this ghost with a physical particle. The price of eliminating the wrong residue sign of the pole, however is an unstable particle as shown below.

So we start from equation \eqref{exp} and enforce the constraint $g_{ab} = \partial_a\, x^{\mu}\, \partial_b\,x_{\mu}$ through a Lagrange multiplier matrix $\gamma^{ab}$ so that the quantum system we actually consider is
\begin{align}
\begin{split}\label{effZ}
    Z & = \int Dx\,Dg\,D\gamma \ \exp{\left(-S[x,g,\gamma]\right)}\,,\\
    S[x,g,\gamma]& = \int d^2{\sigma} \sqrt{g} \ g^{ab} {\cal D}_a x_{\mu }\ \left( T - S\,{\cal D}^2 +\frac{1}{ M^2} {\cal D}^4 \right) \ {\cal D}_b x_{\mu } \\[2mm]
    &\,\,\,\,\,\,\,\,\,\,\,\,\,\,\,\,+ \int d^2{\sigma} \sqrt{g} \, \gamma^{ab}\,(\partial_a\, x^{\mu}\, \partial_b\,x_{\mu} - g_{ab})\ .
    \end{split}
\end{align}
The world-sheet deformations we want to account for are those along the normal directions. In order to do this, we shall parameterize the string world-sheet using a Gauss map such that $x^{\mu} \mapsto (\sigma^0, \sigma^1, \chi^i(\sigma))$ with $\chi^i$ parameterizing the $D-2$ transverse fluctuating modes. Integrating out those modes will allow us to determine an effective potential to one-loop order and derive the gap equations in the saddle-point approximation for constant diagonal elements of the metric
$g_{ab} = \begin{psmallmatrix}
    g_0 & 0 \\
    0 & g_1
\end{psmallmatrix}$
and for a Lagrange multiplier metric of the form $\gamma^{ab} = \gamma\, g^{ab}$ \cite{Diamantini:1998zu}.

It is straightforward to show that the problem of integrating out the transverse modes reduces to a Gaussian integration, which we can do exactly. Thus, integrating out the transverse modes we get the effective action $S_{\text{eff}}=S_{\text{eff}}[\gamma,g_0,g_1]$ given by
\begin{align}\label{effacsp}
    S_{\text{eff}} =& 2A \sqrt{g_0\,g_1} \left[(T + \gamma)\, \frac{g_0 + g_1}{g_0\,g_1} - 2\,\gamma \right] \notag \\[2mm]
    & + 2A \frac{D-2}{(4 \pi)^2}\sqrt{g_0\,g_1} \int d^2k \, \log\left[k^2 \left(T + \gamma -S \, k^2 +  \frac{k^4}{M^2} \right)\right]\,,
\end{align}
where we have introduced a complete set of momentum states to evaluate the determinant, and $A = \int d^2\sigma = \beta L$ denotes the surface extrinsic area where $\beta$ parameterizes the $\sigma^0$-coordinate and $L$ is the length of the string. Since we know that the stiffness $S$ vanishes anyway at the fixed point, as derived above, we will choose $S=0$ henceforth for simplicity of calculation and presentation.

The spectrum is obtained by considering the argument of the $\rm log$ function in \eqref{effacsp} and solving the equation
\begin{equation}
p^2 \left( \left( R^2 + I^2 \right)^2 - 2 \left( R^2-I^2 \right) p^2 + p^4\right)  =0 \ ,
\label{spectrum1}
\end{equation}
where $p$ is the Minkowski space-time two-momentum and
\begin{eqnarray}
R^2 &&= \frac{M} {2}\sqrt{T+\gamma} + \frac{SM^2}{4} \ ,
\nonumber \\
I^2 &&= \frac{M} {2}\sqrt{T+\gamma} - \frac{SM^2}{4} \ .
\label{ri}
\end{eqnarray}
There are two types of excitations: the first are massless phonons satisfying $p^2=0$ as in the Nambu-Goto string, the second is an additional excitation determined by the equation
\begin{equation}
\left( \left( R^2 + I^2 \right)^2 - 2 \left( R^2-I^2 \right) p^2 + p^4\right)  =0 \ .
\label{spectrum2}
\end{equation}
For positive stiffness this gives the spectrum
\begin{equation}
E(k) =\sqrt{(R\mp iI)^2 + k^2} \ ,
\label{spectrum3}
\end{equation}
where $k$ is spatial momentum on the world-sheet and the upper sign propagates forward in time and the lower backwards in time in the Feynman propagator. This describes a resonance with mass $m^2 = SM^2/2$ and half-time life $1/2RI$.

The propagator is determined by the poles of the inverse kinetic operator and is thus given by
\begin{eqnarray}
D(p) &&\propto \frac{1}{p^2 \left( \left( R^2 + I^2 \right)^2 - 2 \left( R^2-I^2 \right) p^2 + p^4\right)}
\nonumber \\
&&= \frac{1}{ (R^2 + I^2)^2}\frac{1}{p^2} - \frac{p^2  - 2(R^2 - I^2)}{(R^2 + I^2)^2\left( \left(p^2 - (R^2 - I^2)\right)^2 + 4 R^2I^2 \right)} \ ,
\label{pfd}
\end{eqnarray}
where the second term is the contribution of the two complex poles in the partial fraction decomposition. The massless pole has positive residue $1/\left(R^2+I^2 \right)^2$. The other two poles are located at $p^2 = (R\pm i I)^2$.

For positive stiffness, the poles lie at positive $p^2=R^2-I^2$. With the usual Breit-Wigner approximation, the spectral density is obtained by setting $p^2 = (R^2-I^2)>0$ in the propagator, which gives
\begin{equation}
\rho_{\rm BW} = \frac{1}{ 4R^2 I^2} \frac{1}{R^2-I^2} \ .
\label{breit}
\end{equation}
Since this is positive for $S>0$ we are dealing with a physical, long-lived particle resonance on the world sheet.

For negative stiffness, the complex conjugate poles move to negative $p^2=(R^2-I^2)<0$. They do not contribute any spectral density on the positive $p^2$-axis anymore. This is characteristic of overdamped intermediate states which are not observable as asymptotic states, analogous to gluons or plasmons.

When interactions are taken into account, the resonances couple to the phonons. The simplest, kinematically allowed coupling is that of a resonance to two phonons. The Brazovskii-Lifschitz resonance is thus an alternative to the later introduced axion for explaining the length-independent branch of string excitations detected in $SU(N)$ lattice computations \cite{Athenodorou:2024vrr, Dubovsky:2013gi}. It is interesting to note that the very broad scattering phase shift observed in the two-phonon channel (see \cite{Athenodorou:2024vrr, Dubovsky:2013gi}) supports the broad, overdamped intermediate state for negative stiffness rather than the sharp particle resonance implied by positive stiffness, yet another indication that confining strings have negative stiffness \cite{Diamantini:1998zu, Diamantini:2002rp}.

In summary, there is a physical excitation on the world-sheet in addition to the Nambu-Goto phonons. For $S>0$ this is a sharp particle resonance, for $S<0$ it is a broad, overdamped mode, not a sharp resonance in the usual sense. The fixed point separates these two regimes, and the lattice comparison above favours the $S<0$, overdamped branch.

\subsection{Saddle-point Effective Potential }

For the one-loop calculation of the effective potential associated with the string action discussed above, we will impose periodic boundary conditions on the ends of the open string. Consequently, we should consider quantized momenta in the direction $\sigma^1$, such that
\begin{align}
    k^2 = (k^0)^2 + (k^1)^2 = (k^0)^2 + \omega_n^2\,, \qquad \omega_n^2 = \frac{\pi^2}{g_1 \,L^2}n^2\,,
\end{align}
In doing the calculations, it is useful to introduce a new mass scale $\mu^2 = M \sqrt{T + \gamma}$ and, after the usual regularization of divergent integrals by analytic continuation \cite{Diamantini:1999uj} we obtain the effective action
\be
\label{speff}
    S_{\text{eff}} = \,2\beta L \sqrt{g_0 g_1} \left[ (T + \gamma)\, \frac{g_0 + g_1}{g_0\,g_1} - 2\,\gamma + \frac{D-2}{2}\, \frac{\mu^2}{8} \right] 
    -\frac{D-2}{2}\frac{\beta \sqrt{g_0}}{4\,\pi} \left[ \frac{\pi^2}{3\,\sqrt{g_1}\,L} +\Omega  \right]\,,
    \ee
where
\be
\Omega := \text{Re}\left[\sum_{n=1}^{\infty} \frac{8 \sqrt{i\,\mu^2}}{n}K_1\left(2\,n\,L\,\sqrt{i\,\mu^2\,g_1} \right) \right]
\ee
and $K_1(z)$ is a modified Bessel function of the second kind. In \cite{Diamantini:1999uj}, all terms that involved the Bessel functions were ignored. Here we will include them and, for the numerical analysis carried out later in this section, we truncate the sum at order $n=10$.

In the saddle point approximation, we find the following gap equations for the modes $\{\gamma,g_0,g_1\}$,

\begin{align}
    &
    \frac{D-2}{2} \frac{\mu^2}{16(T +\gamma)} + \frac{g_0 + g_1}{g_0\,g_1} -2 - \frac{D-2}{2} \frac{1}{8\pi L \sqrt{g_1}} \frac{\partial \Omega}{\partial\gamma} =0\,
    \label{eqA}\\[2mm]
    &\frac{D-2}{2}\left(\frac{\mu^2}{16} - \frac{\pi}{48 L^2 g_1}\right) - \gamma- \frac{D-2}{2}\frac{\Omega}{16\pi L \sqrt{g_1}}  
    +\frac{T + \gamma}{2}\left( \frac{g_0 - g_1}{g_0 g_1} \right) = 0  \,,\label{eqB}\\[2mm]
    & \frac{D-2}{2} \left(\frac{\mu^2}{16} + \frac{\pi}{48 L^2 g_1}\right) - \gamma - \frac{D-2}{2}\frac{\sqrt{g_1}}{8\pi L}\frac{\partial \Omega}{\partial g_1}\,  
    +\frac{T + \gamma}{2}\left( \frac{g_1 - g_0}{g_0 g_1} \right) = 0 \label{eqC}\,.
\end{align}
Substituting back equation \eqref{eqB} into \eqref{speff}, one can show that the effective action \eqref{speff} takes a very simple form
\begin{equation}
    S_{\text{eff}} = 4\,\beta\, L (T + \gamma) \sqrt{\frac{g_1}{g_0}}\,.
\label{simple}
\end{equation}

For simplicity, in the following we shall solve
the gap equation at the fixed point, by setting also $T=0$ and retaining only the dynamical tension $4\gamma$ (see \cite{Diamantini:1998zu}). Summing eqs. \eqref{eqA} and \eqref{eqB} and eqs. \eqref{eqB} and \eqref{eqC} and using the identity
\begin{equation}
\partial_x K_1(x) =-\frac{1}{ x} K_1(x) -K_0(x) \ ,
\label{besselid}
\end{equation}
one obtains a first system of two equations that must be solved simultaneously to determine $\gamma (L)$ and $g_1(L)$. By introducing dimensionless quantities
\be \ell=ML\, \qquad {\rm and} \qquad y=16^2\gamma/M^2
\ee
and selecting:
\be
x_n:= (n/2) \ell y^{1/4}\,\sqrt{i g_1(\ell)}
\ee
we can write these as
\begin{eqnarray}
y&&=y^{1/2}+\frac{8}{ \pi} y^{1/2}  \text{Re}\left[\sum_{n=1}^{\infty} i K_0\left(x_n \right) \right] \ ,
\nonumber \\
y&&=\frac{3}{ 2}\frac{g_1}{ 2g_1-1} y^{1/2} - \frac{16 \pi}{ 3} \frac{1}{ (2g_1-1)\ell^2} +\frac{4}{ \pi} \frac{g_1}{ 2g_1-1} y^{1/2} \text{Re}\left[\sum_{n=1}^{\infty} i K_0\left(x_n\right) \right]
\nonumber \\
&&-\frac{32}{ \pi} \frac{\sqrt{g_1}}{ (2g_1-1) \ell} y^{1/4} \text{Re}\left[\sum_{n=1}^{\infty} \frac{\sqrt{i}}{ n} K_1\left(x_n\right) \right] \ .
\label{geq}
\end{eqnarray}
Finally, from the expression (\ref{simple}) we obtain the interquark effective potential as
$V_{\rm eff} = S_{\rm eff}/\beta$, which, in terms of $y$ and $\ell$ takes the form
\begin{equation}
    V_{\rm eff}(L)=\frac{M^2}{ 64}y(\ell) L \ \sqrt{1-\frac{D-2}{ 2} \frac{128 \pi }{ \ell^2 y(\ell)} \frac{c_{\rm eff} (\ell)}{ 12}} \ .
\label{arvis1}
\end{equation}
Here $c_{\rm eff}(\ell)$ is a running central charge which, as we will see below, is related to the Regge intercept $\alpha_0$ ($\alpha_0(\ell)  = \frac{c_{\text{eff}}(\ell)}{12}$),
\be
 c_{\text{eff}}  = 1 + \frac{6\ell}{\pi^2} \, y^{1/4}(\ell)\,  g_1^{1/2}(\ell)\, \text{Re}\left[ \sum_{n=1}^{\infty} \frac{\sqrt{i}}{n} K_1\left( x_n\right)\right] 
    +\frac{3\ell^2}{2\pi^2} \, y^{1/2}(\ell)\,  g_1(\ell)\, \text{Re}\left[ \sum_{n=1}^{\infty} i\, K_0\left(x_n  \right)\right],
\ee
which at large distances $c_{\text {eff}}\rightarrow 1$ corresponding to the central charge of a Nambu-Goto string mode.

The potential (\ref{arvis1}) is a generalized Arvis potential \cite{Arvis:1983fp} with running parameters that depend on the string size. For sufficiently long strings, it reduces to a Cornell potential
\begin{equation}\label{effV}
    V_{\text{eff}}(L) = \frac{M^2}{64} \,y(\ell) L - \frac{D-2}{2} \frac{\pi}{12\,L}\,c_{\text{eff}}(\ell)\,,
\end{equation}
with fixed parameters dependent on string-size $\ell$\footnote{Note that the effective confinement potentials are usually written in terms of the variable $R$ instead of $L$. We prefer $L$ to emphasize that it is related to the length of an open string.}

In the next section we will present the numerical solutions of these equations.

\subsection{Tension, intercept and quark potential}

If we consider the dominant behaviour of the Brazovskii-Lifshitz string in the limit of infinite string length we obtain a world-sheet conformal field theory with central charge $c=1$ per degree of freedom (see also \cite{Diamantini:1999uj}). This long-string limit can be considered as a ``free string", which represents the reverse of the medal of asymptotic freedom. As soon as we consider shorter strings, though, the finite thickness of the Brazovskii-Lifshitz strings starts having a sizable effect on their behaviour.

As soon as we move away from the IR fixed point, additionally, we start seeing  interactions. These arise from two relevant perturbations encoded in two dimensionless parameters $S$ and $T/M^2$. For simplictiy, in the following we shall consider only the fixed-point strings and we analyze their behaviour when varying the string length.

First of all we remark that the above gap equations can be solved exactly in the long-string limit $L\gg 1/M$. The tension ${\cal T}$ is given by ${\cal T}_\infty = M^2/64$. We shall measure distances in units of $1/M$, by using the dimensionless length $\ell = ML$.

\begin{figure}[!ht]
\centering
	\includegraphics[width=15cm]{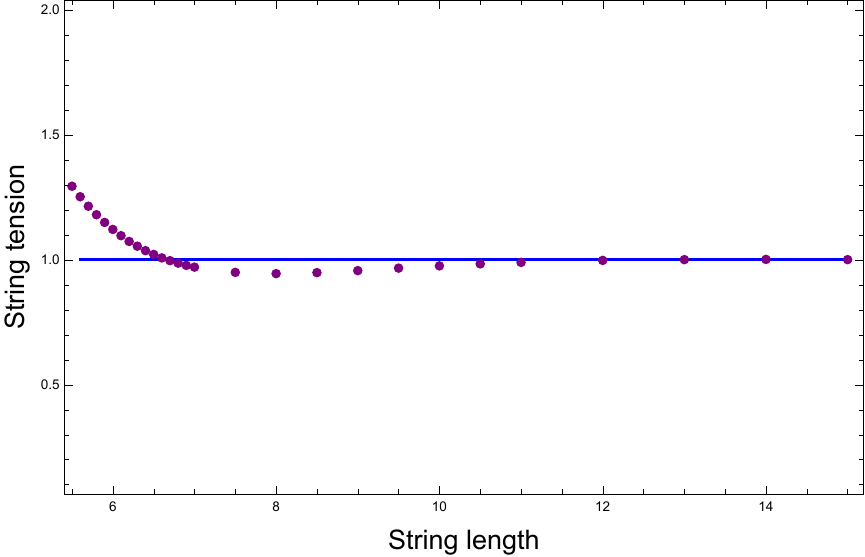}
	\vspace{-0.1cm}
	\caption{String tension in units of ${\cal T}_\infty = m^2/64$ as a function of string length $L$ in units $1/m$.}
	\label{fig:Fig.1}
\end{figure}

The potential (\ref{arvis1}) (and therefore also the string effective action) becomes imaginary for strings of length $\ell < 5.6$. This indicates that the string picture breaks down below a fundamental length scale $O(1/M)$ set by the thickness, as expected. This is the dual of the gauge picture breaking down above scales $1/\Lambda_{QCD}$.
In Fig. \ref{fig:Fig.1} we show the behaviour of the tension $t( \ell)={\cal T}(\ell)/{\cal T}_\infty$ as a function of string length in this string regime.

\begin{figure}[!ht]
\centering
	\includegraphics[width=15cm]{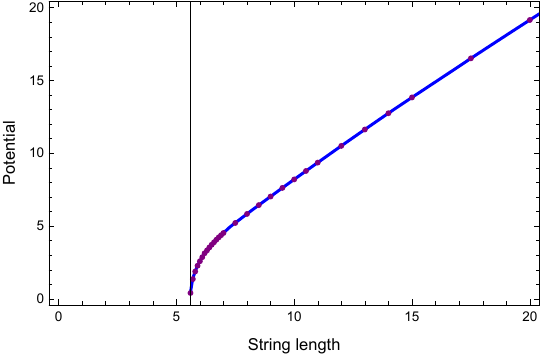}
	\vspace{-0.1cm}
	\caption{The quark-antiquark potential in units of $m/64$.}
	\label{fig:Fig.2}
\end{figure}

The quark potential $V$ (in units of $M/64$) is given by
\begin{equation}
    V_{\rm eff} =y(\ell) \ell \ \sqrt{1-\frac{128 \pi}{ \ell^2 y(\ell)} \alpha_0(\ell)}
    \label{arvis2}
\end{equation}
where $y(\ell)$ and $\alpha_0(\ell)$ are the solutions of the gap equations derived in the previous section. For sufficiently large strings it reduces to
\begin{equation}\label{dimlessV}
    V_{\text{eff}} = y(\ell) \ell - \frac{64\pi}{\ell} \alpha_0(\ell)\,,
\end{equation}
with $\alpha_0(\ell) = c_{\rm eff}(\ell)/12$ and $c_{\rm eff}(\ell)$ is the running central charge, with $c_{\rm eff}(\ell) \to 1$, for $\ell \gg 1$. This potential, shown in Fig. \ref{fig:Fig.2}, automatically emerges as a Arvis-type potential (for a review see \cite{Bali:2000gf}), but with running coefficients dependent on distance. At large distances it takes the form of a Cornell potential with running but fixed coefficients. Note that
this potential is derived, rather than posited.

At large distances the linear term dominates, at short distances the potential becomes steeper. Note however, that this has nothing to do with a gauge regime, it arises purely from confining string properties. We will derive the first few eigenvalues in this potential in the next subsection.

\begin{figure}[!ht]
\centering
	\includegraphics[width=15cm]{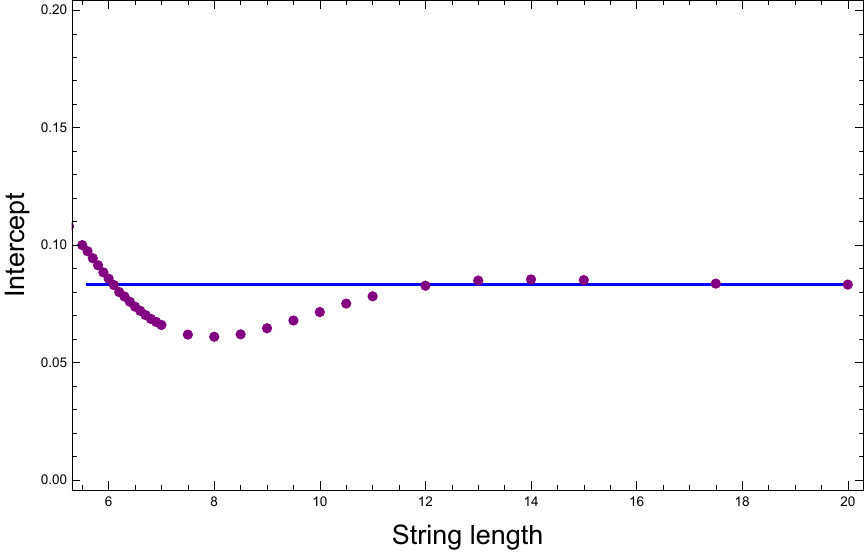}
	\vspace{-0.1cm}
	\caption{The intercept $\alpha_0(mL)$ of the confining string. Shown as a straight line is the Nambu-Goto value $c=1$ value 1/12.}
	\label{fig:Fig.3}
\end{figure}

In string descriptions, the masses $M_{\rm q}$ of high angular momentum $J$ mesons are expected to lie on Regge trajectories (see, e.g. \cite{donnachie2002pomeron})
\begin{eqnarray}
J &&= \alpha^\prime M_{\rm q} ^2+ \alpha_0 \ ,
\nonumber \\
\alpha^\prime && = \frac{1}{ 2\pi {\cal T}} \ .
\label{regge}
\end{eqnarray}
The Nambu-Goto string predicts an intercept $\alpha_0 = C$ where $C$ is the universal quantity $C=(D-2)/24$ corresponding to a central charge $c=1$ per transverse degree of freedom and resulting in an intercept $\alpha_0=1/12$. While the meson masses and angular momenta do lie on a straight line, the intercept $\alpha_0=1/12=0.0833$ predicted by the Nambu-Goto string is too low, the correct value being typically around $\alpha_0 \approx 0.5$. In Fig. \ref{fig:Fig.3} we show the running intercept $\alpha_0(\ell) $ of the Brazovskii-Lifshitz string. At large distances this converges to the universal value 1/12 of the $c=1$ string, as evidenced by the horizontal line. At short distances, however, it shows an increase to higher values.

\subsection{Heavy quarkonium mass splittings}\label{masssp}

Contrary to standard approaches (see, e.g. \cite{Brandt:2016xsp, Brambilla:2014eaa}), in which the quark potential is a phenomenological two-parameter interpolation between a QCD Coulomb potential at short distances and a linear term due to an effective string at large distances, in our result the \emph{functional form} of the quark potential is fixed, with no tunable parameters, since it is derived from a quantum consistent string fixed point. Matching this fixed functional form to the quarkonium spectrum below still requires fixing two numbers, the dimensionless reduced quark mass $\mu_q$ and a Robin boundary parameter $\alpha$ encoding the unknown short-distance completion at the string endpoint (both of which affect the predicted level spacings, and not merely an overall scale), so the comparison below amounts to a two-parameter fit to this fixed functional form rather than a parameter-free prediction. At the IR fixed point, this string, and consequently also the potential, exists if the string length exceeds 5.6 times the fundamental scale $1/M$ representing its width. As we have shown above, there are then two relevant perturbations one can add to this simplest theory.

It is an interesting question, how well the fixed-point potential already predicts the meson spectroscopy. To this end we shall consider the heaviest quarkonium, bottomonium, for which relativistic effects are generally believed to be negligible and which can be studied by solving the Schr\"odinger equation in the derived potential. In particular, we shall focus on the radial states 1S to 6S shown below, \cite{ParticleDataGroup:2022pth}.
\begin{table}[!ht]
\centering
\begin{tabular}{c c}
\hline\hline
State & Mass (MeV) \\
\hline
$\Upsilon(1S)$ & $9460.30 \pm 0.26$ \\
$\Upsilon(2S)$ & $10023.26 \pm 0.31$ \\
$\Upsilon(3S)$ & $10355.2 \pm 0.5$ \\
$\Upsilon(4S)$ & $10579.4 \pm 1.2$ \\
$\Upsilon(5S)$ & $10889.9 \pm 3.2$ \\
$\Upsilon(6S)$ & $10992.9 \pm 3.1$ \\
\hline\hline
\end{tabular}
\caption{PDG world-average masses of the $\Upsilon(nS)$ bottomonium states.}
\label{tab:bottomonium_masses}
\end{table}
Also, as usual, we will focus on the ratios of mass differences
\begin{equation}
r_n = \frac{M(\Upsilon(nS))-M(\Upsilon(1S))}{ M(\Upsilon(2S))-M(\Upsilon(1S))} \ ,
\label{ms}
\end{equation}
for $n=2 \dots 6$. Taking differences eliminates common additive constants independent of the potential, considering ratios eliminates the overall energy scale. This is why these ratios are sensitive to the shape of the potential only. Using the above values we get
\begin{equation}
r_n= (1, 1.59, 1.99, 2.54, 2.72) \ .
\label{massratios}
\end{equation}

We now solve the center of mass radial Schr\"odinger equation for S states of two heavy bottom quarks in a potential $U(r)$ proportional to the one shown in Fig. \ref{fig:Fig.2},
\begin{equation}
\left( -\frac{1}{ M_q} \frac{d^2 }{ d r^2} + U(r) \right) u(r) = E\  u(r) \ ,
\label{sch1}
\end{equation}
in natural units, with $u(r) = r R(r)$ and $M_q$ the quark mass. We now introduce again the dimensionless length $x=Mr$ and we absorb the overall scale of the potential into a dimensionless mass parameter $\mu_q$ and dimensionless energies $e$ measured in units of a scale $\propto M$,
\begin{equation}
\left( -\frac{1}{ \mu_q} \frac{d^2}{ d x^2} + V(x) \right) u(x)  = e u(x) \ .
\label{sch2}
\end{equation}

In solving this equation numerically we use an order 3 spline interpolation of the string potential between $x_{\rm min}=5.6$, the smallest size of the string, and $x=30$, and a linear potential for $x\ge 30$, where it coincides with the asymptotic exact solution.
Of course, we have to impose a boundary condition on wave functions at the onset of the string regime $x_{\rm min} = 5.6$. To avoid imposing an artificial hard wall, we adopt Robin boundary conditions,
\begin{equation}
u^{\prime}\left( x_{\rm min} \right) - \alpha u\left( x_{\rm min} \right) = 0 \ .
\label{robin}
\end{equation}
These provide the most general self-adjoint realization of the Hamiltonian at the onset of the string regime. The Robin parameter $\alpha$ determines the character of the wall at the boundary as can be seen by expanding the wave function $u$ there,
\begin{equation}
     u\left( x \right) = u\left( x_{\rm min} \right) + u^{\prime} \left( x_{\rm min} \right) \left( x-x_{\rm min} \right) = u\left( x_{\rm min} \right) \left( 1+ \alpha\left(x- x_{\rm min} \right) \right) \ .
    \label{robinint}
\end{equation}
This shows that, although the probability is all concentrated in the physical domain, the boundary conditions ``simulate" a wave function that vanishes at $x= x_{\rm min} -1/\alpha$. A negative $\alpha$ represents an absorbing wall, $\alpha=0$ is a perfectly reflecting wall corresponding to von Neumann boundary conditions and a positive $\alpha$ corresponds to a repulsive wall, with $\alpha \to \infty$ encoding Dirichlet boundary conditions. We also impose Dirichlet boundary conditions $u(x) = 0$ at $x= 30$. This upper boundary is chosen to lie well inside the classically forbidden region, where the wave function is already exponentially suppressed (we have checked that the total probability in the region $x \in(25,30)$ never exceeds $10^{-27}$).

\begin{figure}[!ht]
\centering
	\includegraphics[width=15cm]{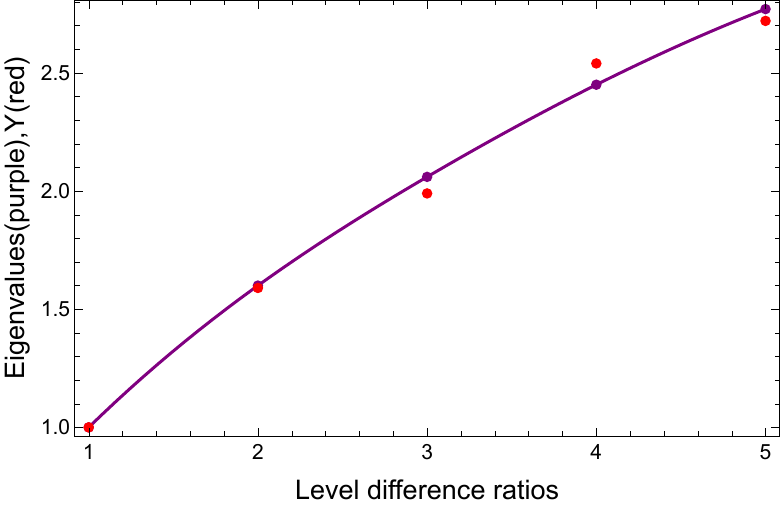}
	\caption{The fit of eigenvalue difference ratios to mass difference ratios for Upsilon radial excitations.}
	\label{fig:Fig.4}
\end{figure}

We compute the 6 lowest eigenvalues of the so-defined Schr\"odinger operator and we form the same ratios of differences $e_n$ to compare with (\ref{ms}) and (\ref{massratios}) by varying the mass parameter $\mu_q$ in the kinetic termand the Robin parameter $\alpha$. The optimal values
\begin{equation}
e_n = (1, 1.6, 2.06, 2.45, 2.77) \ ,
\label{fitb}
\end{equation}
with a total standard deviation of 2.5\% with respect to the experimental ones, are shown in Fig. \ref{fig:Fig.4}.

\begin{figure}[!ht]
\centering
	\includegraphics[width=15cm]{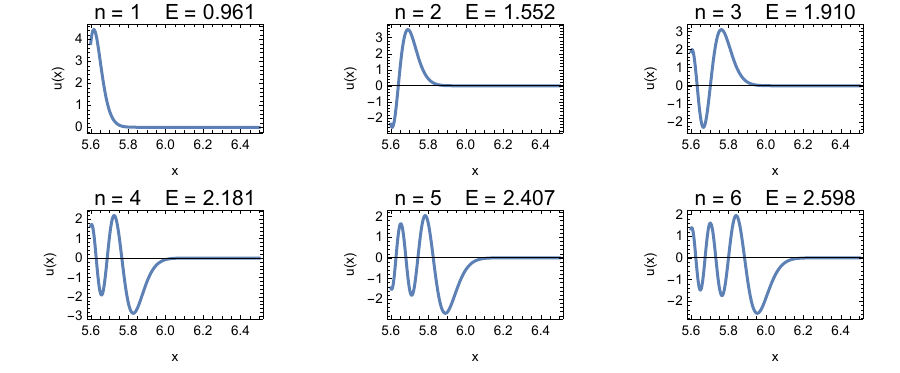}
	\vspace{-0.1cm}
	\caption{The wave functions of the first 6 states in the confining string potential, shown as a function of dimensionless lengths $x=rM$.}
	\label{fig:Fig.5}
\end{figure}

It is worth highlighting that these optimal fits are obtained for a value $\alpha =21.8 $ of the Robin parameter, which is very near to Dirichlet boundary conditions, as explained above. The wave functions of these optimal states are automatically localized in the initial stringy region of the potential,
as evidenced in Fig. \ref{fig:Fig.5}.

Our result lends strong support to the exclusive stringy origin of the $\Upsilon$ states, with no influence of physics below the string scale $x_{\rm min}$, especially in view of the fact that this is just a fixed-point result, with no inclusion of any relevant perturbation around it. This result also bears out Polyakov's original conjecture of the universality of confinement.

\section{Conclusions}\label{sec6}

We have investigated the description of confinement in terms of a UV-complete confining string arising from a gauge theory directly in (3+1) dimensions. Starting from compact QED in $(3+1)$ dimensions with a $\theta$–term, we showed that, in the dyon condensation phase, the theory admits a dual formulation in terms of a massive two-form field $B_{\mu\nu}$ leading to a string theory that, when removing the cutoff at strong coupling along the double-scaling trajectory of section \ref{sec4}, flows to a Brazovskii–Lifshitz infrared fixed point. This fixed point describes  a UV-complete (3+1)-dimensional string with a finite thickness, arising from dimensional transmutation whose long-distance behaviour approaches that of the Nambu–Goto string. But, contrary to the Nambu-Goto string, this confining string can be consistently defined a UV-complete theory in $3+1$ dimensions. The non-vanishing $\theta$ plays a crucial role in making the two-dimensional theory quantum mechanically consistent in the sense that the cutoff can be removed leading to finite couplings.

Our main results are:
\begin{itemize}
\item
{ IR string fixed point:} compact QED with a $\theta$–term flows, along the double-scaling trajectory identified in section \ref{sec4} (whose relation to the generic continuum limit of the confining phase of compact QED remains an open question), to a Brazovskii–Lifshitz fixed point describing a confining string with an intrinsic thickness.
\item{Additional world-sheet excitation:} Brazovskii-Lifshitz strings contain a world-sheet excitation -- a sharp resonance for positive stiffness, a broad, overdamped mode for negative stiffness -- that can be considered as an alternative to the worldsheet axion \cite{Dubovsky:2013gi, Athenodorou:2024vrr} to improve fitting with the lattice data. Lattice data support negative stiffness also for $SU(N)$, i.e.\ the broad, overdamped regime rather than a sharp resonance.
\item
{ Quark potential:} at one-loop the theory predicts a generalized Arvis \cite{Arvis:1983fp} potential $V(L)= a(L)L\sqrt{1-2b(L)/a(L)L^2}$ with running coefficients entirely determined by the string dynamics.
\item{ Regge intercept:} finite thickness tends to raise the intercept from the Nambu–Goto value $\alpha_0=1/12$ to higher values.
\item { Phenomenology:} solving the Schr\"odinger equation for the heaviest quarkonium states $\Upsilon(nS)$, with the reduced quark mass and a single boundary-condition parameter fixed by a two-parameter fit to the fixed functional form of the potential, yields mass–splitting ratios in $2.5\% $ agreement with experiment and point to the exclusive stringy origin of these states.
\end{itemize}

In close analogy with critical strings, which couple naturally to a massless two-form, the confining strings described here couple to a massive antisymmetric tensor field $B_{\mu\nu}$. The mass introduces the physical scale associated with the thickness of the confining string.

Dualities among massive antisymmetric tensor theories play a central role in our analysis. In particular,  compact QED in four dimensions, after dyon condensation, can be formulated in terms of a massive antisymmetric tensor field $B_{\mu\nu}$.

Recent developments in generalized and non-invertible symmetries touch on similar properties of antisymmetric tensor symmetries and may also be important for a more complete formulation of the dynamics of confining strings \cite{Gaiotto:2014kfa}. Likewise, $T{\bar T}$ and $J{\bar T}$ deformations could play a role in a full description of the effective two-dimensional theory (see for instance \cite{Dubovsky:2017cnj}). These directions provide promising avenues for future work. Also, here we have mainly studied the physics of the IR fixed point. Further investigations of the consequences of a non-zero stiffness for QCD will be the subject of forthcoming publications.

Overall, our results provide concrete support for Polyakov’s conjecture that confining gauge theories admit a universal infrared string description. In this picture the Brazovskii–Lifshitz string represents, in some way, the infrared dual of asymptotically free gauge dynamics, capturing the universal long-distance behaviour of confining flux tubes independently of the microscopic gauge group. Beyond supporting the string picture of confinement, this framework opens several directions for further investigation, including the role of relevant perturbations around the fixed point, the extension to dynamical quarks, and detailed comparisons with lattice results for flux-tube spectra and potentials. We hope to address some of these questions in the near future.

\section*{Acknowledgements}
We thank Cliff Burgess, Shanta de Alwis, Sebastian Cespedes, Sergei Dubovsky, Christopher Thomas and Gonzalo Villa for useful communications. MCD and CAT thank New York University Abu Dhabi (NYUAD) for hospitality at different stages of this project. FQ and LFZ thank the administration and security office of NYUAD, as well as the UAE government, for their great effort in keeping them safe during the last stages of this project. They are funded by Tamkeen under the research grant ADHPG-AD457 to NYUAD.

\bibliographystyle{JHEP.bst}
\bibliography{ref}

@book{Greensite2020,
  title     = {An Introduction to the Confinement Problem},
  author    = {Greensite, Jeff},
  series    = {Lecture Notes in Physics},
  volume    = {821},
  year      = {2020},
  edition   = {2nd},
  publisher = {Springer},
  address   = {Cham, Switzerland},
  isbn      = {978-3-030-51562-1},
  doi       = {10.1007/978-3-030-51562-1}
}

@article{Gross:2022hyw,
    author = "Gross, Franz and others",
    title = "{50 Years of Quantum Chromodynamics}",
    eprint = "2212.11107",
    archivePrefix = "arXiv",
    primaryClass = "hep-ph",
    doi = "10.1140/epjc/s10052-023-11949-2",
    journal = "Eur. Phys. J. C",
    volume = "83",
    pages = "1125",
    year = "2023"
}

@article{tHooft:1977nqb,
    author = "'t Hooft, Gerard",
    title = "{On the Phase Transition Towards Permanent Quark Confinement}",
    reportNumber = "Print-78-0099 (UTRECHT)",
    doi = "10.1016/0550-3213(78)90153-0",
    journal = "Nucl. Phys. B",
    volume = "138",
    pages = "1--25",
    year = "1978"
}

@inproceedings{tHooft:1999cgx,
    author = "'t Hooft, Gerard",
    title = "{Monopoles, instantons and confinement}",
    booktitle = "{National Summer School for Graduate Students: We-Heraeus Doktorandenschule Saalburg: Grundlagen und Neue Methoden der Theoretischen Physik}",
    eprint = "hep-th/0010225",
    archivePrefix = "arXiv",
    month = "9",
    year = "1999"
}

@article{Mandelstam:1974pi,
    author = "Mandelstam, S.",
    title = "{Vortices and Quark Confinement in Nonabelian Gauge Theories}",
    reportNumber = "PRINT-74-1623 (UC,BERKELEY)",
    doi = "10.1016/0370-1573(76)90043-0",
    journal = "Phys. Rept.",
    volume = "23",
    pages = "245--249",
    year = "1976"
}

@article{DiGiacomo:1999yas,
    author = "Di Giacomo, A. and Lucini, B. and Montesi, L. and Paffuti, G.",
    title = "{Color confinement and dual superconductivity of the vacuum. 1.}",
    eprint = "hep-lat/9906024",
    archivePrefix = "arXiv",
    reportNumber = "IFUP-TH-35-99",
    doi = "10.1103/PhysRevD.61.034503",
    journal = "Phys. Rev. D",
    volume = "61",
    pages = "034503",
    year = "2000"
}

@article{DiGiacomo:1999fb,
    author = "Di Giacomo, A. and Lucini, B. and Montesi, L. and Paffuti, G.",
    title = "{Color confinement and dual superconductivity of the vacuum. 2.}",
    eprint = "hep-lat/9906025",
    archivePrefix = "arXiv",
    reportNumber = "IFUP-TH-36-99",
    doi = "10.1103/PhysRevD.61.034504",
    journal = "Phys. Rev. D",
    volume = "61",
    pages = "034504",
    year = "2000"
}

@article{Carmona:2001ja,
    author = "Carmona, J. M. and D'Elia, Massimo and Di Giacomo, A. and Lucini, B. and Paffuti, G.",
    title = "{Color confinement and dual superconductivity of the vacuum. 3.}",
    eprint = "hep-lat/0103005",
    archivePrefix = "arXiv",
    reportNumber = "IFUP-TH-2001-10",
    doi = "10.1103/PhysRevD.64.114507",
    journal = "Phys. Rev. D",
    volume = "64",
    pages = "114507",
    year = "2001"
}

@article{Luscher:2002qv,
    author = "Luscher, Martin and Weisz, Peter",
    title = "{Quark confinement and the bosonic string}",
    eprint = "hep-lat/0207003",
    archivePrefix = "arXiv",
    reportNumber = "CERN-TH-2002-138, MPI-PHT-2002-24",
    doi = "10.1088/1126-6708/2002/07/049",
    journal = "JHEP",
    volume = "07",
    pages = "049",
    year = "2002"
}

@article{Juge:2002br,
    author = "Juge, K. Jimmy and Kuti, Julius and Morningstar, Colin",
    title = "{Fine structure of the QCD string spectrum}",
    eprint = "hep-lat/0207004",
    archivePrefix = "arXiv",
    doi = "10.1103/PhysRevLett.90.161601",
    journal = "Phys. Rev. Lett.",
    volume = "90",
    pages = "161601",
    year = "2003"
}

@article{HariDass:2007tx,
    author = "Hari Dass, N. D. and Majumdar, Pushan",
    title = "{Continuum limit of string formation in 3-d SU(2) LGT}",
    eprint = "hep-lat/0702019",
    archivePrefix = "arXiv",
    reportNumber = "MS-TP-07-2",
    doi = "10.1016/j.physletb.2007.08.097",
    journal = "Phys. Lett. B",
    volume = "658",
    pages = "273--278",
    year = "2008"
}

@article{Caselle:2016mqu,
    author = "Caselle, Michele and Panero, Marco and Vadacchino, Davide",
    title = "{Width of the flux tube in compact U(1) gauge theory in three dimensions}",
    eprint = "1601.07455",
    archivePrefix = "arXiv",
    primaryClass = "hep-lat",
    doi = "10.1007/JHEP02(2016)180",
    journal = "JHEP",
    volume = "02",
    pages = "180",
    year = "2016"
}

@article{Aharony:2013ipa,
    author = "Aharony, Ofer and Komargodski, Zohar",
    title = "{The Effective Theory of Long Strings}",
    eprint = "1302.6257",
    archivePrefix = "arXiv",
    primaryClass = "hep-th",
    reportNumber = "WIS-01-13-FEB-DPPA",
    doi = "10.1007/JHEP05(2013)118",
    journal = "JHEP",
    volume = "05",
    pages = "118",
    year = "2013"
}

@article{Paulos:2016but,
    author = "Paulos, Miguel F. and Penedones, Joao and Toledo, Jonathan and van Rees, Balt C. and Vieira, Pedro",
    title = "{The S-matrix bootstrap II: two dimensional amplitudes}",
    eprint = "1607.06110",
    archivePrefix = "arXiv",
    primaryClass = "hep-th",
    reportNumber = "CERN-TH-2016-163",
    doi = "10.1007/JHEP11(2017)143",
    journal = "JHEP",
    volume = "11",
    pages = "143",
    year = "2017"
}

@article{EliasMiro:2019kyf,
    author = "Elias Mir{\'o}, Joan and Guerrieri, Andrea L. and Hebbar, Aditya and Penedones, Jo{\~a}o and Vieira, Pedro",
    title = "{Flux Tube S-matrix Bootstrap}",
    eprint = "1906.08098",
    archivePrefix = "arXiv",
    primaryClass = "hep-th",
    doi = "10.1103/PhysRevLett.123.221602",
    journal = "Phys. Rev. Lett.",
    volume = "123",
    number = "22",
    pages = "221602",
    year = "2019"
}

@article{EliasMiro:2021nul,
    author = "Elias Mir{\'o}, Joan and Guerrieri, Andrea",
    title = "{Dual EFT bootstrap: QCD flux tubes}",
    eprint = "2106.07957",
    archivePrefix = "arXiv",
    primaryClass = "hep-th",
    doi = "10.1007/JHEP10(2021)126",
    journal = "JHEP",
    volume = "10",
    pages = "126",
    year = "2021"
}

@article{Athenodorou:2024loq,
    author = "Athenodorou, Andreas and Dubovsky, Sergei and Luo, Conghuan and Teper, Michael",
    title = "{Confining strings and the worldsheet axion from the lattice}",
    eprint = "2411.03435",
    archivePrefix = "arXiv",
    primaryClass = "hep-th",
    doi = "10.1142/S0217751X25400032",
    journal = "Int. J. Mod. Phys. A",
    volume = "40",
    number = "33",
    pages = "2540003",
    year = "2025"
}

@article{Diamantini:1995yb,
    author = "Diamantini, M. C. and Sodano, P. and Trugenberger, C. A.",
    title = "{Gauge theories of Josephson junction arrays}",
    eprint = "hep-th/9511168",
    archivePrefix = "arXiv",
    reportNumber = "CERN-TH-95-294, DFUPG-111-95, UGVA-DPT-1995-11-907",
    doi = "10.1016/0550-3213(96)00309-4",
    journal = "Nucl. Phys. B",
    volume = "474",
    pages = "641--677",
    year = "1996"
}

@article{Diamantini:2018mjg,
    author = "Diamantini, M. C. and Trugenberger, C. A. and Vinokur, V. M.",
    title = "{Confinement and Asymptotic Freedom with Cooper pairs}",
    eprint = "1807.01984",
    archivePrefix = "arXiv",
    primaryClass = "hep-th",
    journal = "APS Physics",
    volume = "1",
    pages = "77",
    year = "2018"
}

@article{Athenodorou:2011rx,
    author = "Athenodorou, Andreas and Bringoltz, Barak and Teper, Michael",
    title = "{Closed flux tubes and their string description in D=2+1 SU(N) gauge theories}",
    eprint = "1103.5854",
    archivePrefix = "arXiv",
    primaryClass = "hep-lat",
    reportNumber = "DESY-11-045",
    doi = "10.1007/JHEP05(2011)042",
    journal = "JHEP",
    volume = "05",
    pages = "042",
    year = "2011"
}

@article{Witten:1998zw,
  author        = {Edward Witten},
  title         = {Anti-de Sitter Space, Thermal Phase Transition, and Confinement in Gauge Theories},
  journal       = {Adv. Theor. Math. Phys.},
  volume        = {2},
  pages         = {505--532},
  year          = {1998},
  doi           = {10.4310/ATMP.1998.v2.n3.a3},
  eprint        = {hep-th/9803131},
  archivePrefix = {arXiv}
}

@article{Maldacena:1997re,
    author = "Maldacena, Juan Martin",
    title = "{The Large N limit of superconformal field theories and supergravity}",
    eprint = "hep-th/9711200",
    archivePrefix = "arXiv",
    journal = "Adv. Theor. Math. Phys.",
    volume = "2",
    pages = "231--252",
    year = "1998",
    doi = "10.4310/ATMP.1998.v2.n2.a1"
}

@article{Gubser:1998bc,
    author = "Gubser, S. S. and Klebanov, I. R. and Polyakov, Alexander M.",
    title = "{Gauge theory correlators from noncritical string theory}",
    eprint = "hep-th/9802109",
    archivePrefix = "arXiv",
    journal = "Phys. Lett. B",
    volume = "428",
    pages = "105--114",
    year = "1998",
    doi = "10.1016/S0370-2693(98)00377-3"
}

@article{Witten:1998qj,
    author = "Witten, Edward",
    title = "{Anti-de Sitter space and holography}",
    eprint = "hep-th/9802150",
    archivePrefix = "arXiv",
    journal = "Adv. Theor. Math. Phys.",
    volume = "2",
    pages = "253--291",
    year = "1998",
    doi = "10.4310/ATMP.1998.v2.n2.a2"
}

@article{Aharony:2010db,
  author        = {Ofer Aharony and Nizan Klinghoffer},
  title         = {Corrections to Nambu-Goto Energy Levels from the Effective String Action},
  journal       = {JHEP},
  volume        = {12},
  pages         = {058},
  year          = {2010},
  doi           = {10.1007/JHEP12(2010)058},
  eprint        = {1008.2648},
  archivePrefix = {arXiv}
}

@article{Teper:2009uf,
    author = "Teper, Michael",
    editor = "Praszalowicz, Michal",
    title = "{Large N and confining flux tubes as strings - a view from the lattice}",
    eprint = "0912.3339",
    archivePrefix = "arXiv",
    primaryClass = "hep-lat",
    journal = "Acta Phys. Polon. B",
    volume = "40",
    pages = "3249--3320",
    year = "2009"
}

@incollection{diamantini2023superinsulation,
  author    = {Diamantini, M. Cristina and Trugenberger, Carlo A. and Vinokur, Valerii M.},
  title     = {Superinsulation},
  booktitle = {Encyclopedia of Condensed Matter Physics},
  edition   = {2nd},
  publisher = {Academic Press},
  address   = {Cambridge, USA},
  pages     = {804--816},
  year      = {2023},
  doi       = {10.1016/B978-0-323-90800-9.00220-1},
  isbn      = {9780323914086}
}

@article{Kleinert:1987,
  author  = {H. Kleinert},
  title   = {Spontaneous Generation of String Tension and Quark Potential},
  journal = {Physical Review Letters},
  volume  = {58},
  number  = {19},
  pages   = {1915--1918},
  year    = {1987},
  doi     = {10.1103/PhysRevLett.58.1915}
}

@article{Kleinert:1996,
  author        = {Hagen Kleinert and Alexander M. Chervyakov},
  title         = {Evidence for Negative Stiffness of QCD Strings},
  journal       = {Phys. Lett. B},
  volume        = {381},
  pages         = {286--292},
  year          = {1996},
  doi           = {10.1016/0370-2693(96)00509-8},
  eprint        = {hep-th/9601030},
  archivePrefix = {arXiv}
}

@article{Polyakov:1975rs,
    author = "Polyakov, Alexander M.",
    editor = "Taylor, J. C.",
    title = "{Compact Gauge Fields and the Infrared Catastrophe}",
    doi = "10.1016/0370-2693(75)90162-8",
    journal = "Phys. Lett. B",
    volume = "59",
    pages = "82--84",
    year = "1975"
}

@article{Aharony:2024ctf,
    author = "Aharony, Ofer and Barel, Netanel and Sheaffer, Tal",
    title = "{Effective strings in QED$_{3}$}",
    eprint = "2412.01313",
    archivePrefix = "arXiv",
    primaryClass = "hep-th",
    doi = "10.1007/JHEP03(2025)143",
    journal = "JHEP",
    volume = "03",
    pages = "143",
    year = "2025"
}

@article{Caselle:2024zoh,
    author = "Caselle, Michele and Magnoli, Nicodemo and Nada, Alessandro and Panero, Marco and Panfalone, Dario and Verzichelli, Lorenzo",
    title = "{Confining strings in three-dimensional gauge theories beyond the Nambu-Got{\={o}} approximation}",
    eprint = "2407.10678",
    archivePrefix = "arXiv",
    primaryClass = "hep-lat",
    doi = "10.1007/JHEP08(2024)198",
    journal = "JHEP",
    volume = "08",
    pages = "198",
    year = "2024"
}

@article{Brandt:2016xsp,
    author = "Brandt, Bastian B. and Meineri, Marco",
    title = "{Effective string description of confining flux tubes}",
    eprint = "1603.06969",
    archivePrefix = "arXiv",
    primaryClass = "hep-th",
    doi = "10.1142/S0217751X16430016",
    journal = "Int. J. Mod. Phys. A",
    volume = "31",
    number = "22",
    pages = "1643001",
    year = "2016"
}

@article{Athenodorou:2024vrr,
    author = "Athenodorou, Andreas and Dubovsky, Sergei and Luo, Conghuan and Teper, Michael",
    title = "{Towards an Effective String Theory for the flux-tube}",
    eprint = "2411.16507",
    archivePrefix = "arXiv",
    primaryClass = "hep-lat",
    doi = "10.22323/1.466.0393",
    journal = "PoS",
    volume = "LATTICE2024",
    pages = "393",
    year = "2025"
}

@article{Polyakov:1997tj,
    author = "Polyakov, Alexander M.",
    editor = "Bais, F. A. and Bergshoeff, E. A. and de Wit, B. and Dijkgraaf, R. and Schellekens, A. N. and Verlinde, Erik P. and Verlinde, Herman L.",
    title = "{String theory and quark confinement}",
    eprint = "hep-th/9711002",
    archivePrefix = "arXiv",
    reportNumber = "PUPT-1743",
    doi = "10.1016/S0920-5632(98)00135-2",
    journal = "Nucl. Phys. B Proc. Suppl.",
    volume = "68",
    pages = "1--8",
    year = "1998"
}

@article{Luscher:1980fr,
    author = "Luscher, M. and Symanzik, K. and Weisz, P.",
    title = "{Anomalies of the Free Loop Wave Equation in the WKB Approximation}",
    reportNumber = "DESY-80-31",
    doi = "10.1016/0550-3213(80)90009-7",
    journal = "Nucl. Phys. B",
    volume = "173",
    pages = "365",
    year = "1980"
}

@article{Luscher:1980iy,
    author = "Luscher, M. and Munster, G. and Weisz, P.",
    title = "{How Thick Are Chromoelectric Flux Tubes?}",
    reportNumber = "DESY-80-63",
    doi = "10.1016/0550-3213(81)90151-6",
    journal = "Nucl. Phys. B",
    volume = "180",
    pages = "1--12",
    year = "1981"
}

@article{Polchinski:1991ax,
    author = "Polchinski, Joseph and Strominger, Andrew",
    title = "{Effective string theory}",
    reportNumber = "UTTG-17-91",
    doi = "10.1103/PhysRevLett.67.1681",
    journal = "Phys. Rev. Lett.",
    volume = "67",
    pages = "1681--1684",
    year = "1991"
}

@article{Aharony:2009gg,
    author = "Aharony, Ofer and Karzbrun, Eyal",
    title = "{On the effective action of confining strings}",
    eprint = "0903.1927",
    archivePrefix = "arXiv",
    primaryClass = "hep-th",
    reportNumber = "WIS-04-09-MAR-DPP",
    doi = "10.1088/1126-6708/2009/06/012",
    journal = "JHEP",
    volume = "06",
    pages = "012",
    year = "2009"
}

@article{Aharony:2010cx,
    author = "Aharony, Ofer and Field, Matan",
    title = "{On the effective theory of long open strings}",
    eprint = "1008.2636",
    archivePrefix = "arXiv",
    primaryClass = "hep-th",
    reportNumber = "WIS-10-10-AUG-DPPA",
    doi = "10.1007/JHEP01(2011)065",
    journal = "JHEP",
    volume = "01",
    pages = "065",
    year = "2011"
}

@article{Dubovsky:2012sh,
    author = "Dubovsky, Sergei and Flauger, Raphael and Gorbenko, Victor",
    title = "{Effective String Theory Revisited}",
    eprint = "1203.1054",
    archivePrefix = "arXiv",
    primaryClass = "hep-th",
    doi = "10.1007/JHEP09(2012)044",
    journal = "JHEP",
    volume = "09",
    pages = "044",
    year = "2012"
}

@article{Dubovsky:2015zey,
    author = "Dubovsky, Sergei and Gorbenko, Victor",
    title = "{Towards a Theory of the QCD String}",
    eprint = "1511.01908",
    archivePrefix = "arXiv",
    primaryClass = "hep-th",
    doi = "10.1007/JHEP02(2016)022",
    journal = "JHEP",
    volume = "02",
    pages = "022",
    year = "2016"
}

@article{Burgess:2025geh,
    author = "Burgess, C. P. and Quevedo, F.",
    title = "{Dual is Different: EFTs, Axions and Nonpropagating Form Fields in Cosmology}",
    eprint = "2509.11340",
    archivePrefix = "arXiv",
    primaryClass = "hep-th",
    month = "9",
    year = "2025"
}

@article{Gabai:2025hwf,
    author = "Gabai, Barak and Gorbenko, Victor and Qiao, Jiaxin",
    title = "{Yang-Mills flux tube in AdS}",
    eprint = "2508.08250",
    archivePrefix = "arXiv",
    primaryClass = "hep-th",
    doi = "10.1007/JHEP06(2026)029",
    journal = "JHEP",
    volume = "06",
    pages = "029",
    year = "2026"
}

@article{Gabai:2026myo,
    author = "Gabai, Barak and Gorbenko, Victor and Offertaler, Bendeguz",
    title = "{Yang-Mills Flux Tube in AdS II: Effective String Theory}",
    eprint = "2602.16694",
    archivePrefix = "arXiv",
    primaryClass = "hep-th",
    month = "2",
    year = "2026"
}

@article{Diamantini:1998zu,
    author = "Diamantini, M. C. and Kleinert, H. and Trugenberger, C. A.",
    title = "{Strings with negative stiffness and hyperfine structure}",
    eprint = "hep-th/9810171",
    archivePrefix = "arXiv",
    doi = "10.1103/PhysRevLett.82.267",
    journal = "Phys. Rev. Lett.",
    volume = "82",
    pages = "267--270",
    year = "1999"
}

@article{Diamantini:1999uj,
    author = "Diamantini, M. Christina and Kleinert, Hagen and Trugenberger, C. A.",
    title = "{Universality class of confining strings}",
    eprint = "hep-th/9903208",
    archivePrefix = "arXiv",
    doi = "10.1016/S0370-2693(99)00532-8",
    journal = "Phys. Lett. B",
    volume = "457",
    pages = "87--93",
    year = "1999"
}

@article{lifshitz1942theory,
  author   = {Lifshitz, Evgeny Mikhailovich},
  title    = {On the Theory of Phase Transitions of the Second Order I},
  journal  = {Journal of Physics (USSR)},
  volume   = {6},
  number   = {1--2},
  pages    = {61--74},
  year     = {1942}
}

@article{Hornreich:1975zz,
    author = "Hornreich, R. M. and Luban, Marshall and Shtrikman, S.",
    title = "{Critical Behavior at the Onset of k --{\ensuremath{>}} -Space Instability on the lamda Line}",
    doi = "10.1103/PhysRevLett.35.1678",
    journal = "Phys. Rev. Lett.",
    volume = "35",
    pages = "1678--1681",
    year = "1975"
}

@article{brazovskii1975phase,
  author  = {Brazovskii, Samuel A.},
  title   = {Phase transition of an isotropic system to a nonuniform state},
  journal = {Soviet Physics JETP},
  volume  = {41},
  number  = {1},
  pages   = {85--89},
  year    = {1975}
}

@article{Sonnenschein:2016pim,
    author = {Sonnenschein, Jacob},
    title = {Holography Inspired Stringy Hadrons},
    eprint = {1602.00704},
    archivePrefix = {arXiv},
    primaryClass = {hep-th},
    doi = {10.1016/j.ppnp.2016.06.005},
    journal = {Prog. Part. Nucl. Phys.},
    volume = {92},
    pages = {1--49},
    year = {2017}
}

@article{Quevedo:1996uu,
    author = "Quevedo, Fernando and Trugenberger, Carlo A.",
    title = "{Phases of antisymmetric tensor field theories}",
    eprint = "hep-th/9604196",
    archivePrefix = "arXiv",
    reportNumber = "CERN-TH-96-109, UGVA-DPT-1996-04-923",
    doi = "10.1016/S0550-3213(97)00337-4",
    journal = "Nucl. Phys. B",
    volume = "501",
    pages = "143--172",
    year = "1997"
}

@article{Polyakov:1996nc,
    author = "Polyakov, Alexander M.",
    title = "{Confining strings}",
    eprint = "hep-th/9607049",
    archivePrefix = "arXiv",
    reportNumber = "PUPT-1632",
    doi = "10.1016/S0550-3213(96)00601-3",
    journal = "Nucl. Phys. B",
    volume = "486",
    pages = "23--33",
    year = "1997"
}

@article{Diamantini:1996vf,
    author = "Diamantini, M. C. and Quevedo, F. and Trugenberger, C. A.",
    title = "{Confining string with topological term}",
    eprint = "hep-th/9612103",
    archivePrefix = "arXiv",
    reportNumber = "CERN-TH-96-319, UGVA-DPT-1996-10-995",
    doi = "10.1016/S0370-2693(97)00132-9",
    journal = "Phys. Lett. B",
    volume = "396",
    pages = "115--121",
    year = "1997"
}

@article{Alvarez:1981kc,
    author = "Alvarez, Orlando",
    title = "{The Static Potential in String Models}",
    reportNumber = "CLNS-81/486",
    doi = "10.1103/PhysRevD.24.440",
    journal = "Phys. Rev. D",
    volume = "24",
    pages = "440",
    year = "1981"
}

@article{Arvis:1983fp,
    author = "Arvis, J. F.",
    title = "{The Exact $q \bar{q}$ Potential in Nambu String Theory}",
    reportNumber = "LPTENS 83/6",
    doi = "10.1016/0370-2693(83)91640-4",
    journal = "Phys. Lett. B",
    volume = "127",
    pages = "106--108",
    year = "1983"
}

@article{Dubovsky:2013gi,
    author = "Dubovsky, Sergei and Flauger, Raphael and Gorbenko, Victor",
    title = "{Evidence from Lattice Data for a New Particle on the Worldsheet of the QCD Flux Tube}",
    eprint = "1301.2325",
    archivePrefix = "arXiv",
    primaryClass = "hep-th",
    doi = "10.1103/PhysRevLett.111.062006",
    journal = "Phys. Rev. Lett.",
    volume = "111",
    number = "6",
    pages = "062006",
    year = "2013"
}

@book{polchinski2005string,
  title={String Theory},
  author={Polchinski, J.},
  isbn={9780521672290},
  series={Cambridge monographs on mathematical physics},
  url={https://books.google.ae/books?id=sXsuYAAACAAJ},
  year={2005},
  publisher={Cambridge University Press}
}

@article{Polyakov:1986cs,
    author = "Polyakov, Alexander M.",
    title = "{Fine Structure of Strings}",
    doi = "10.1016/0550-3213(86)90162-8",
    journal = "Nucl. Phys. B",
    volume = "268",
    pages = "406--412",
    year = "1986"
}

@article{Kleinert:1986bk,
    author = "Kleinert, H.",
    title = "{The Membrane Properties of Condensing Strings}",
    doi = "10.1016/0370-2693(86)91111-1",
    journal = "Phys. Lett. B",
    volume = "174",
    pages = "335--338",
    year = "1986"
}

@article{Witten:1979ey,
    author = "Witten, Edward",
    title = "{Dyons of Charge e theta/2 pi}",
    reportNumber = "CERN-TH-2724",
    doi = "10.1016/0370-2693(79)90838-4",
    journal = "Phys. Lett. B",
    volume = "86",
    pages = "283--287",
    year = "1979"
}

@article{Diamantini:1997wk,
    author = "Diamantini, M. C. and Trugenberger, Carlo A.",
    title = "{Surfaces with long range correlations from noncritical strings}",
    eprint = "hep-th/9712008",
    archivePrefix = "arXiv",
    reportNumber = "UGVA-DPT-1997-11-992",
    doi = "10.1016/S0370-2693(98)00023-9",
    journal = "Phys. Lett. B",
    volume = "421",
    pages = "196--202",
    year = "1998"
}

@book{Polyakov:1987hqn,
    author = "Polyakov, A. M.",
    title = "{Gauge Fields and Strings}",
    doi = "10.1201/9780203755082",
    isbn = "978-1-351-44609-9, 978-3-7186-0393-0, 978-0-203-75508-2",
    publisher = "Taylor {\&} Francis",
    address = "London",
    year = "1987"
}

@article{Diamantini:2002rp,
    author = "Diamantini, M. C. and Trugenberger, C. A.",
    title = "{QCD like behaviour of high temperature confining strings}",
    eprint = "hep-th/0202178",
    archivePrefix = "arXiv",
    doi = "10.1103/PhysRevLett.88.251601",
    journal = "Phys. Rev. Lett.",
    volume = "88",
    pages = "251601",
    year = "2002"
}

@article{Brambilla:2014eaa,
    author = "Brambilla, Nora and Groher, Michael and Martinez, Hector E. and Vairo, Antonio",
    title = "{Effective string theory and the long-range relativistic corrections to the quark-antiquark potential}",
    eprint = "1407.7761",
    archivePrefix = "arXiv",
    primaryClass = "hep-ph",
    reportNumber = "TUM-EFT-36-12",
    doi = "10.1103/PhysRevD.90.114032",
    journal = "Phys. Rev. D",
    volume = "90",
    number = "11",
    pages = "114032",
    year = "2014"
}

@article{ParticleDataGroup:2022pth,
    author = "Workman, R. L. and others",
    collaboration = "Particle Data Group",
    title = "{Review of Particle Physics}",
    doi = "10.1093/ptep/ptac097",
    journal = "PTEP",
    volume = "2022",
    pages = "083C01",
    year = "2022"
}

@article{Bali:2000gf,
    author = "Bali, Gunnar S.",
    title = "{QCD forces and heavy quark bound states}",
    eprint = "hep-ph/0001312",
    archivePrefix = "arXiv",
    reportNumber = "HUB-EP-99-67",
    doi = "10.1016/S0370-1573(00)00079-X",
    journal = "Phys. Rept.",
    volume = "343",
    pages = "1--136",
    year = "2001"
}

@book{donnachie2002pomeron,
  title     = {Pomeron Physics and QCD},
  author    = {Donnachie, Sandy and Dosch, Günther and Landshoff, Peter and Nachtmann, Otto},
  year      = {2002},
  publisher = {Cambridge University Press},
  series    = {Cambridge Monographs on Particle Physics, Nuclear Physics, and Cosmology},
  number    = {19},
  isbn      = {9780521780391}
}

@article{Gaiotto:2014kfa,
    author = "Gaiotto, Davide and Kapustin, Anton and Seiberg, Nathan and Willett, Brian",
    title = "{Generalized Global Symmetries}",
    eprint = "1412.5148",
    archivePrefix = "arXiv",
    primaryClass = "hep-th",
    doi = "10.1007/JHEP02(2015)172",
    journal = "JHEP",
    volume = "02",
    pages = "172",
    year = "2015"
}

@article{Dubovsky:2017cnj,
    author = "Dubovsky, Sergei and Gorbenko, Victor and Mirbabayi, Mehrdad",
    title = "{Asymptotic fragility, near AdS$_{2}$ holography and $ T\overline{T} $}",
    eprint = "1706.06604",
    archivePrefix = "arXiv",
    primaryClass = "hep-th",
    doi = "10.1007/JHEP09(2017)136",
    journal = "JHEP",
    volume = "09",
    pages = "136",
    year = "2017"
}

\end{document}